\documentclass[a4paper,10pt]{article}
\pdfoutput=1
\usepackage{jheppub}
\usepackage{dsfont}
\usepackage{graphicx}
\usepackage{amsmath}
\usepackage{amsfonts}
\usepackage{amssymb}
\usepackage{amsthm}
\usepackage{mathtools}
\usepackage{empheq}
\usepackage{subcaption}
\usepackage{mathtools}
\usepackage[shortlabels]{enumitem}
\usepackage{physics}
\usepackage{mathrsfs}
\usepackage{slashed}
\usepackage{cellspace}
\usepackage{dsfont}
\usepackage{tikz}
\usetikzlibrary{arrows,shapes}
\usetikzlibrary{trees}
\usetikzlibrary{patterns}
\usetikzlibrary{matrix,arrows}
\usetikzlibrary{positioning}
\usetikzlibrary{calc,through}
\usetikzlibrary{decorations.pathreplacing}
\usepackage{pgffor}	
\usetikzlibrary{decorations.pathmorphing}
\usetikzlibrary{decorations.markings}
\usepackage{todonotes}
\usepackage{slashed}
\tikzset{
    vector/.style={decorate, decoration={snake}, draw},
	provector/.style={decorate, decoration={snake,amplitude=2.5pt}, draw},
	antivector/.style={decorate, decoration={snake,amplitude=-2.5pt}, draw},
        smallvector/.style={decorate, decoration={snake,amplitude=1.5pt,post length=0.5mm}, draw},
    fermion/.style={draw=black, postaction={decorate},
        decoration={markings,mark=at position .55 with {\arrow[draw=black]{>}}}},
    fermionbar/.style={draw=black, postaction={decorate},
        decoration={markings,mark=at position .55 with {\arrow[draw=black]{<}}}},
    fermionnoarrow/.style={draw=black},
    gluon/.style={decorate, draw=black,
        decoration={coil,amplitude=4pt, segment length=5pt}},
    scalar/.style={dashed,draw=black, postaction={decorate},
        decoration={markings,mark=at position .55 with {\arrow[draw=black]{>}}}},
    scalarbar/.style={dashed,draw=black, postaction={decorate},
        decoration={markings,mark=at position .55 with {\arrow[draw=black]{<}}}},
    scalarnoarrow/.style={dashed,draw=black},
    electron/.style={draw=black, postaction={decorate},
        decoration={markings,mark=at position .55 with {\arrow[draw=black]{>}}}},
    bigvector/.style={decorate, decoration={snake,amplitude=4pt}, draw},
    arrow/.style={draw=black, postaction={decorate},
        decoration={markings,mark=at position 1 with {\arrow[draw=black]{>}}}},
}

    \newcommand{\cb}[1]{
\begin{tikzpicture}[#1]
\draw (-0.2ex,0ex) -- (1.2ex,0ex);
\draw (0.8ex,0ex) -- (0.2ex,-1ex);
\draw (-0.2ex,-1ex) -- (1.2ex,-1ex);
\draw (0.8ex,-1ex) -- (0.2ex,0ex);
\end{tikzpicture}
}

\newcommand{\nb}[1]{
\begin{tikzpicture}[#1]
\draw (-0.2ex,0) -- (1.2ex,0);
\draw (0.8ex,0) -- (0.8ex,-1ex);
\draw (-0.2ex,-1ex) -- (1.2ex,-1ex);
\draw (0.2ex,-1ex) -- (0.2ex,0);
\end{tikzpicture}
}

\newcommand{\tric}[1]{
\begin{tikzpicture}[#1]
\draw (-0.2ex,0) -- (1.2ex,0);
\draw (0.2ex,0) -- (0.5ex,-1ex);
\draw (0.8ex,0) -- (0.5ex,-1ex);
\draw (-0.2ex,-1ex) -- (1.2ex,-1ex);
\end{tikzpicture}
}

\newcommand{\tricinv}[1]{
\begin{tikzpicture}[#1]
\draw (-0.2ex,0) -- (1.2ex,0);
\draw (0.2ex,-1ex) -- (0.5ex,0ex);
\draw (0.8ex,-1ex) -- (0.5ex,0ex);
\draw (-0.2ex,-1ex) -- (1.2ex,-1ex);
\end{tikzpicture}
}

\newcommand{\bub}[1]{
\begin{tikzpicture}
\draw (-0.2ex,0) -- (1.2ex,0);
\draw (-0.2ex,-1ex) -- (1.2ex,-1ex);
\draw (0.5ex,0) arc (40:-40:0.8ex);
\draw (0.5ex,0) arc (140:220:0.8ex);
\end{tikzpicture}
}

\tikzstyle{block} = [draw, rectangle, 
    minimum height=3em, minimum width=6earticlem]

    \usepackage{tikz-feynman}
\tikzfeynmanset{compat=1.1.0}
\tikzfeynmanset{
graviton/.style={circle, draw=green!60, fill=green!5, very thick, minimum size=7mm}
}
\tikzfeynmanset{
codot/.style={/tikz/shape=circle,
/tikz/fill=red,/tikz/minimum size=0.1cm,/tikz/inner sep=1.8pt}
}
\tikzfeynmanset{sb/.style=
{/tikz/shape=circle,
/tikz/fill=black,
 /tikz/minimum size=0.3cm,/tikz/inner sep=1.pt
 } }
\tikzfeynmanset{myblob/.style=
{/tikz/shape=ellipse,
/tikz/fill=red,
 /tikz/minimum width=0.5cm,
 } }
 \tikzfeynmanset{myblobFF/.style=
{/tikz/shape=circle,
 /tikz/minimum width=1cm,
 } }
 \tikzfeynmanset{myblob2/.style=
{/tikz/shape=rectangle,
/tikz/fill=red,
 /tikz/minimum width=0.1cm,/tikz/inner sep=1.8pt
 } }
  \tikzfeynmanset{myvertex/.style=
{/tikz/shape=circle,
/tikz/fill=black,
 /tikz/minimum width=0.05cm,
 } }
 \tikzfeynmanset{GR/.style={/tikz/shape=ellipse,/tikz/fill={rgb:black,1;white,2},/tikz/minimum size=0.3cm,} }
\tikzset{box/.pic={\filldraw[fill=black]  (0,0) circle (2.5pt);
				   \filldraw [fill=black] (0.5,0) circle (2.5pt);
			       \draw [line width=5pt] (0,0) -- (0.5,0);}}

\tikzset{wiggle/.style={decorate, decoration=snake}}

 \tikzfeynmanset{HV/.style=
{/tikz/shape=circle,
/tikz/fill={rgb:black,1;white,2},
 /tikz/minimum size=0.01cm,/tikz/inner sep=1.8pt
 } }

\def\oh{\mathcal{O}}

\begin{document}

\title{Bound States in Perturbative Quantum Gravity with Hydrogen-like Degeneracy}
\author[a]{Callum R. T. Jones,}
\emailAdd{crtjones@arizona.edu}
\author[b]{Shruti Paranjape,}
\emailAdd{shruti\_paranjape@brown.edu}
\author[b]{Marcos Skowronek}
\emailAdd{marcos\_skowronek\_santos@brown.edu}
\affiliation[a]{
Department of Physics, University of Arizona, Tucson, AZ 85721, USA
}
\affiliation[b]{
Department of Physics, Brown University, Providence, RI 02912, USA
}

\abstract{Recent results in the study of gravitational scattering amplitudes indicate that some highly-symmetric relativistic systems may exactly conserve a version of the Laplace-Runge-Lenz (LRL) vector in two-body bound states. We make a systematic study, in the context of a generic EFT of long-range interactions due to the exchange of massless mediator particles of spins 0, $\frac{1}{2}$, 1, $\frac{3}{2}$, 2, of the conditions for the conservation of a hidden LRL vector; both classically (no orbital precession) and quantum-mechanically (hydrogen-like degeneracy of bound states). The calculations require several new technical developments including the extension of relativistic post-Minkowskian potential matching to quantum corrections and the incorporation of long-range forces due to the exchange of pairs of massless fermions. We find that while classically the absence of precession is a generic property of a large class of models, including Kaluza-Klein theories, with vector and scalar exchanges, maintaining the degeneracy quantum-mechanically requires a surprising cancellation between gravitons and 6 Majorana gravitinos, hinting at a special role for $\mathcal{N}=6$ supergravity. 
}

\maketitle
\flushbottom

\section{Introduction}
\label{sec:intro}

There has been tremendous recent progress in the calculation of both bound and unbound observables in the post-Minkowskian (PM) perturbative expansion of classical gravitational two-body systems using methods from on-shell scattering amplitudes \cite{Cheung:2018wkq,Bern:2019nnu,Bern:2021yeh,Dlapa:2021npj,Dlapa:2021vgp,Driesse:2024xad,Driesse:2024feo,Bern:2025wyd,Driesse:2026qiz}. This is of course motivated by the need for improved precision predictions for gravitational wave observables in upcoming third generation ground and space-based detectors \cite{Reitze:2019iox,Punturo:2010zz,LISA:2017pwj,ET:2025xjr}. In addition to this phenomenological motivation, these developments have led to new theoretical insights, such as analytic continuations between bound and unbound observables \cite{Kalin:2019rwq,Kalin:2019inp,Cho:2021arx,Khalaf:2023ozy}, have brought about new formulations of quantum field theories and methods of calculation \cite{Kosower:2018adc,Mogull:2020sak,Caron-Huot:2023vxl,DiVecchia:2023frv,Kim:2025gis} and revealed surprising simplicity in highly symmetric idealized systems \cite{Caron-Huot:2018ape,Parra-Martinez:2020dzs,Guevara:2023wlr,Bern:2025zno}. 

For obvious reasons, most of these calculations are restricted to classical observables. The \textit{quantum-first} pipeline that has been developed, which begins with quantum scattering amplitudes and then subsequently expands in the $\hbar\rightarrow 0$ limit recovering classical observables as expectation values, has some unique advantages over traditional methods \cite{Westpfahl:1985tsl,Goldberger:2004jt}. The combination of unitarity methods \cite{Bern:1994zx,Bern:1994cg,Bern:1997sc} together with the double-copy construction of gravitational amplitudes from gauge theory counterparts \cite{Kawai:1985xq,Bern:2008qj,Bern:2010ue} gives an efficient means of taming the complex nonlinearities of general relativity. Moreover, by working directly with relativistic scattering amplitudes we retain manifest Lorentz invariance, allowing for the efficient calculation of all-orders-in-velocity observables packaging together an infinite number of post-Newtonian (PN) contributions. 

Intuitively, if the low-energy EFT of quantum gravity is under calculational control such that it allows for the efficient extraction of classical observables then it should also be useful for calculating the leading (semi-classical) quantum corrections. Calculations of this kind in the non-relativistic limit have a long history; in some cases, such as the famous calculation of the leading quantum correction to Newton's law \cite{Donoghue:1993eb,Donoghue:1994dn,BjerrumBohr:2002kt,Bjerrum-Bohr:2013bxa}, the quantum corrections are universal low-energy predictions of quantum gravity as they do not depend on the Wilson coefficients characterizing the UV completion. So far, relatively little work has been done synthesizing these quantum gravitational scattering calculations with recent PM relativistic methods (for some exceptions to this see \cite{Bjerrum-Bohr:2021din,DiVecchia:2023frv}); one of the goals of this paper is a first attempt to make such a connection.

As an application, we revisit an example of a surprising hidden property of some special two-body bound systems from on-shell scattering amplitudes: the conservation of a relativistic and quantum analogue of the Laplace-Runge-Lenz (LRL) vector \cite{Burgess:2003qv,Caron-Huot:2014gia,Davis:2023zqv,deNeeling:2023egt,deNeeling:2024vbz,Guevara:2024edh}. Recall that for two-body systems with a Coulomb-like potential
\begin{equation}
    \label{eq:CoulombH}
    H^{(0)}(\mathbf{x},\mathbf{p}) = \frac{\mathbf{p}^2}{2m} - \frac{\alpha}{r}\,,
\end{equation}
we can define a vector $\mathbf{A} = \mathbf{p}\times \mathbf{L} - m\alpha\hat{\mathbf{r}}$ that is conserved under time evolution $\dot{\mathbf{A}} = 0$. This triplet of additional conserved charges combines with the $SO(3)$ rotation generators $\mathbf{L}$ giving an enlarged $SO(4)$ Poisson algebra that manifests as the miraculous integrability of Keplerian orbits \cite{Goldstein1,Goldstein2}. In particular, the existence of the conserved LRL vector implies that the orbits form closed ellipses with a periapsis that does not precess. Moreover the conservation of the LRL vector persists in the quantized model described by the Hamiltonian (\ref{eq:CoulombH}); as shown first by Pauli \cite{Pauli:1926qpj}, the quantum $SO(4)$ symmetry completely fixes the spectrum of bound states in hydrogenic systems and implies the peculiar $l$-independent  degeneracy of energy levels
\begin{equation}
    \label{eq:H0energies}
    E_{n,l}^{(0)} = -\frac{m\alpha^2}{2n^2}\,.
\end{equation}
Relativistic corrections, such as those that occur in classical general relativity \cite{Einstein:1938yz}, or from the Dirac equation \cite{Dirac:1928hu,Darwin,Gordon1928} explicitly break the LRL conservation, leading to PN orbital precession and atomic fine structure. In the less physical spinless case, the energy levels found by solving the Klein-Gordon equation with a Coulomb potential were derived by Fock \cite{Fock1926}\footnote{Equivalently, this closed-form expression re-sums the perturbations to the energy levels from replacing the non-relativistic kinetic energy in (\ref{eq:CoulombH}) with the fully relativistic expression $\sqrt{\mathbf{p}^2 +m^2}$. These kinetic energy corrections are \textit{universal} in the sense that they are common to all Hamiltonians obtained by taking the non-relativistic limit of a relativistic system, and so the LRL symmetry is \textit{generically broken} unless finely tuned interactions in the effective potential cancel the breaking of the degeneracy.}
\begin{align}
    \label{eq:KGenergies}
    E^{\text{KG}}_{n,l} &= m\left[1+\frac{\alpha^2}{\left(n-l-\frac{1}{2}+\sqrt{\left(l+\frac{1}{2}\right)^2-\alpha^2}\right)^2}\right]^{-1/2}\nonumber\\
    &= m -\frac{m\alpha^2}{2n^2} -\frac{m\alpha^4}{2n^3}\left(\frac{1}{l+\frac{1}{2}}-\frac{3}{4n}\right) + \mathcal{O}\left(\alpha^6\right).
\end{align}
 Beginning at $\mathcal{O}(\alpha^4)$ the result depends explicitly on the orbital quantum number $l$, showing that relativistic corrections generically break the $SO(4)$ LRL symmetry. 

In an intriguing paper, Caron-Huot and Zahraee analyzed a classical two-body bound system of $\frac{1}{2}$-BPS black holes in $\mathcal{N}=8$ supergravity using scattering amplitudes \cite{Caron-Huot:2018ape}. They found that the 2PM correction to the orbital precession \textit{vanishes} as a consequence of the so-called \textit{no-triangle} property of the corresponding scattering amplitudes \cite{Arkani-Hamed:2008owk}. This observation was interpreted as evidence that this highly-special system preserves an analogue of the LRL vector even when relativistic corrections are included. The authors of \cite{Caron-Huot:2018ape} conjectured that this extends to quantum corrections to the system, a precise version of this conjecture being that the special interactions of this model produce an effective potential that \textit{cancels} the universal relativistic kinetic energy corrections in (\ref{eq:KGenergies}), restoring an $l$-independent hydrogen-like spectrum of bound states. Note that a similar property was previously proven to hold for quantum bound states of W-bosons on the Coulomb branch of planar $\mathcal{N}=4$ super-Yang-Mills \cite{Caron-Huot:2014gia}; identified as a highly-non-obvious consequence of dual conformal invariance.

In this paper we explicitly test this conjecture and show that it is \textit{false}. We develop a quantum-mechanical version of the PM potential matching introduced in \cite{Cheung:2018wkq} and apply it to a completely general EFT of massless mediators. This includes both bosonic and fermionic mediators up to spin-2. Our most interesting observation concerns the competing contributions of spin-$2$ gravitons and spin-$3/2$ gravitinos. We argue that for the degeneracy to be preserved in large angular momentum bound states, the 2PM quantum correction to the potential must exactly vanish. In an ultra-relativistic expansion of the potential ($\sigma\rightarrow \infty$), only gravitons and gravitinos give a contribution at leading-order and can be made to exactly cancel only for a model containing \textit{exactly} 6 Majorana gravitinos. This remarkable observation is found to be consistent with previous calculations of Regge scattering in $\mathcal{N}$-extended (massless) supergravity \cite{Bartels:2012ra,Bern:2020gjj} and implies that only a model of $\mathcal{N}=6$ supergravity can possibly preserve an exact form of the LRL vector. We are unable to identify a physical model for which the cancellation extends to the complete 2PM contribution, but given the limitations of our EFT construction we cannot definitively rule it out either. It therefore is left as a tantalizing conjecture: the true (gravitational) hydrogen atom of the 21$^\text{st}$ century may in fact be found in $\mathcal{N}=6$ supergravity! 

The paper is organized as follows. In Section \ref{sec:review}, we discuss which classical and quantum observables demonstrate the LRL symmetry of a two-body system and how to derive them from the classical radial action and quantum potential, respectively. Section \ref{sec:eft_ops} contains a list of EFT operators that we consider in the most generic examples. In Section \ref{sec:classical}, we present the classical condition for no-precession. We highlight two examples: BPS-$\overline{\text{BPS}}$ bound states in Section \ref{sec:BPS_KK} and $\mathcal{N}\ge 4$ supergravity in Section \ref{sec:Nge4}. Next in Section \ref{sec:quantum}, we present the main result of the paper: the quantum conditions for bound-state degeneracy. We discuss the inclusion of fermions in Section \ref{sec:fermion} and why $\mathcal{N}=6$ supergravity is special in Section \ref{subsec:gravitino}. Finally, we show how LRL symmetry is \emph{not} conserved beyond the classical limit in $\mathcal{N}=8$ supergravity. Novel features of our results and future directions are discussed in Section \ref{sec:discussion}.

\paragraph{Notation and conventions:} Throughout this paper, unless otherwise indicated, we use natural units $\hbar = c = 1$. Lorentz covariant expressions are defined with the mostly minus metric signature $\eta_{\mu\nu} = \eta^{\mu\nu} = \text{diag}\left(+1,-1,-1,-1\right)$. Loop integrals are evaluated using dimensional regularization, in $D=4-2\epsilon$ dimensions. We suppress factors of $2\pi$ using: $\hat{\text{d}}^{D}\ell := {\text{d}^{D}\ell}/{(2 \pi)^D}$ and $\hat{\delta}^{(D)}(...):=(2\pi)^D \delta^{(D)}(...)$. Perturbations of the metric around the Minkowski background are defined as $g_{\mu\nu} = \eta_{\mu\nu} + \kappa h_{\mu\nu}$, where $\kappa := \sqrt{32\pi G}$.

\section{Post-Minkowskian Scattering and Bound States}
\label{sec:review}
In this section, we describe how scattering amplitudes in the Post-Minkowskian expansion can be used to extract observables that indicate the presence of LRL symmetry in a two-body system. In Section \ref{subsec:reviewclassical}, we present a review of the construction of the long-range part of the 2PM scattering amplitude using unitarity methods and the calculation of the classical orbital precession from the classical radial action. This is followed in Section \ref{subsec:reviewquantum} by a discussion of the quantum potential and its bound-state degeneracy, another indicator of LRL symmetry. Finally in Section \ref{sec:eft_ops}, we present the interaction content of the EFT that we study in Sections \ref{sec:classical} and \ref{sec:quantum}. 

\subsection{Classical radial action and orbital precession}
\label{subsec:reviewclassical}
In this paper we study two-body, classical and quantum, bound systems via the relativistic scattering amplitude for a pair of massive non-spinning states $\Phi_1$ and $\Phi_2$ with masses $m_1$ and $m_2$ respectively
\begin{equation}
    \label{eq:firstamp} \mathcal{M}_4\bigg((\Phi_1)_{p_1} + (\Phi_2)_{p_2} \rightarrow (\Phi_1)_{p_1+q} + (\Phi_2)_{p_2-q}\bigg).
\end{equation}
We work in the so-called \textit{post-Minkowskian} (PM) regime of scattering, which is defined by a hierarchy of scales where the angular momentum of the two-body system is large compared to Planck's constant: $J := |\mathbf{p}||\mathbf{b}| \gg\hbar$ \cite{Cheung:2018wkq,Bern:2019crd,Kosower:2018adc}, where $\mathbf{p}$ is the initial relative 3-momentum in the center-of-mass (COM) frame and $\mathbf{b}$ the relative impact parameter. When this expansion is applied to appropriate semi-classical observables, the leading-order term corresponds to the classical limit and higher-orders to quantum corrections. Prior to the Fourier transform to impact parameter space, the PM expansion can be organized in powers of the soft momentum transferred between the bodies
\begin{equation}
    \label{eq:softexpansion}
    q^\mu \ll p_i^\mu \sim m_i\,,
\end{equation}
where the interaction is mediated by the exchange of massless force-carrying particles. Loop integral contributions are expanded prior to integration using the method of regions \cite{Beneke:1997zp}. The integrals we encounter have two regions: a \textit{soft region} $\ell^\mu \sim q^\mu$ that contains non-analytic in $q^2$ terms (around $q^2=0$) and a \textit{hard region} $\ell^\mu \sim p_i^\mu$ that contains analytic in $q^2$ terms. After Fourier transforming, the hard region gives only short-range delta function contributions, the long-range interactions are therefore given entirely by the soft region. 

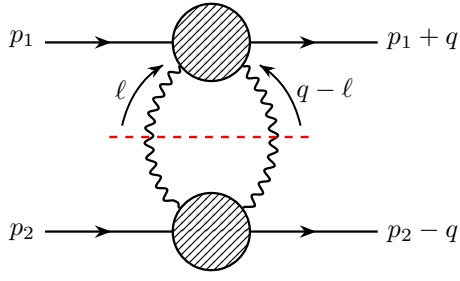
\begin{figure}[t]

\centering

\begin{tikzpicture}[thick, >=Stealth, midarrow/.style={postaction={decorate,decoration={markings,mark=at position 0.52 with {\arrow{Stealth[length=2.2mm]}}}}}, fermion/.style={line width=0.85pt, midarrow}, scalar/.style={decorate, decoration={snake, amplitude=1.6pt, segment length=5.5pt}, line width=0.85pt}, curvedmom/.style={line width=0.75pt, postaction={decorate,decoration={markings,mark=at position 1 with {\arrow{Stealth[length=2mm]}}}}}, cut/.style={red, dashed, line width=0.9pt}, blob/.style={circle, draw=black, line width=0.85pt, minimum size=1.02cm, pattern=north east lines, pattern color=black}]

\coordinate (T) at (0,1.25);

\coordinate (B) at (0,-1.25);

\coordinate (TL) at (-0.42,0.94);

\coordinate (TR) at (0.42,0.94);

\coordinate (BL) at (-0.42,-0.94);

\coordinate (BR) at (0.42,-0.94);

\node[blob] at (T) {};

\node[blob] at (B) {};

\draw[fermion] (-2.20,1.25) -- (-0.51,1.25);

\draw[fermion] (0.51,1.25) -- (2.20,1.25);

\draw[fermion] (-2.20,-1.25) -- (-0.51,-1.25);

\draw[fermion] (0.51,-1.25) -- (2.20,-1.25);

\draw[scalar] (TL) .. controls (-0.92,0.35) and (-0.92,-0.35) .. (BL);

\draw[scalar] (TR) .. controls (0.92,0.35) and (0.92,-0.35) .. (BR);

\draw[curvedmom] (-1.18,0.15) .. controls (-1.10,0.50) and (-0.92,0.78) .. (-0.64,0.95);

\draw[curvedmom] (1.18,0.15) .. controls (1.10,0.50) and (0.92,0.78) .. (0.64,0.95);

\draw[cut] (-1.35,0) -- (1.35,0);

\node[left] at (-2.20,1.3) {$p_1$};

\node[right] at (2.20,1.3) {$p_1+q$};

\node[left] at (-2.20,-1.25) {$p_2$};

\node[right] at (2.20,-1.25) {$p_2-q$};

\node at (-1.20,0.62) {$\ell$};

\node at (1.50,0.62) {$q-\ell$};

\end{tikzpicture}
\caption{Massless $t$-channel cut of the 1-loop amplitude, contributing to the long-range interaction between the massive bodies.}

\label{fig:t-channel-cut}

\end{figure}

For gravitational scattering, the $L$-loop contribution to the amplitude of order $\mathcal{O}\left(G^{L{+}1}q^{L{-}2}\right)$ captures the $\mathcal{O}\left(\hbar^{0}\right)$ classical dynamics and gives a long-range contribution to the effective potential of the form $V(r)\sim r^{-1-L}$ in $D=4$ spacetime dimensions. Similarly, the $\oh(G^{L+1}q^{L-1})$ part of the amplitude gives the $\oh(\hbar)$ quantum correction, scaling as $V(r)\sim r^{-2-L}$. At 1-loop the classical non-analytic terms have the form $(-q^2)^{-1/2}$ and the quantum non-analytic terms behave as $\text{log}(-q^2)$. Both such terms have branch cuts in the complex $t:=-q^2$-plane and can therefore be reconstructed from their discontinuity, which is in turn fixed by the $t$-channel unitarity cut depicted in Figure \ref{fig:t-channel-cut}.\footnote{As always there is a delicate question of unitarity cut reconstruction in $D=4$ vs. $D=4-2\epsilon$. At 1-loop, the error introduced by using the former is a \textit{rational} contribution to the amplitude. Since the only available kinematic invariants are $\sigma := \frac{p_1\cdot p_2}{m_1 m_2}$ and $q^2$, there are no possible rational functions that scale as $q^{-1}$ and so there is no ambiguity at classical order. For quantum terms, rational contributions can in principle appear and would scale as $q^0$ in the soft expansion. However, since this is analytic in $q^2$ it does not give a contribution to the long-range potential, and so we lose no information by sewing 4-dimensional tree-level amplitudes.} The input required for the calculation of the classical and quantum contributions to the long-range potential at 1-loop are the tree-level Compton amplitudes involving two massive states and two massless mediators. Explicitly, we need the following cut
\begin{equation}
\label{eq:comptoncut}
\begin{aligned}
    \text{Cut}_{\bub{}}\left[\mathcal{M}_4^{\text{1-loop}}\right] := \sum_{\text{states}}\mathcal{M}_4^{\text{tree}}\left((\overline{\Phi}_1)_{-p_1}, \overline{\varphi}_{\ell_1}, \varphi_{-\ell_2}, (\Phi_1)_{p_1+q}\right)\mathcal{M}_4^{\text{tree}}\left((\overline{\Phi}_2)_{-p_2}, \overline{\varphi}_{\ell_2}, \varphi_{-\ell_1}, (\Phi_2)_{p_2-q}\right),
\end{aligned}
\end{equation}
where $\varphi,\overline{\varphi}$ are massless mediators. We have chosen a convention for the Compton amplitudes in which all the momenta are outgoing and by definition the right-hand-side is evaluated on the cut kinematics $\ell^2 = (\ell - q)^2 =0$. The long-range part of the integrand is then given by writing the unitarity cut in a manifestly local and covariant form and then promoting it away from the cut conditions and reintroducing the cut propagators; more information on the unitarity method can be found in the original references \cite{Bern:1994zx,Bern:1994cg,Bern:1997sc}. A full list of the scalar integrals that appear in this work and the details of their soft expansion are reviewed in Appendix \ref{app:integrals}.

One of our main objectives in this paper is to characterize when a given EFT has LRL symmetry. Classically, this is equivalent to the condition that orbits do not precess. In this work, different classical observables will be extracted from the quantum scattering amplitude via its relation to the radial action \cite{Bern:2021dqo,Damgaard:2021ipf,Kol:2021jjc, Damgaard:2023ttc}. In the classical limit, the amplitude is known to exponentiate in impact parameter space
\begin{equation}\label{eq:radial action exp}
    i\mathcal{M}(q) = 4 \frac{m_1 m_2 \sqrt{\sigma^2-1}}{\hbar^2} \int \text{d}^2 b\ e^{-\frac{i}{\hbar}q\cdot b} \left(e^{i I_r(J)}-1\right),
\end{equation}
where $I_r(J)$ denotes the classical radial action of the system and $\sigma = \frac{p_1\cdot p_2}{m_1 m_2}$. Performing a perturbative expansion in powers of $G$
\begin{equation}
    \mathcal{M}(q) = \sum_{n=1}^\infty \mathcal{M}^{(n)}(q)\,,\qquad I_r(J) = \sum_{n=1}^\infty I_r^{(n)}(J)\,,
\end{equation}
we can identify the different terms order by order; at tree- and 1-loop-level,
\begin{align}
    \mathcal{M}^{(1)}(q) &=\frac{1}{\hbar}\tilde I_r^{(1)}(q),\\
    \mathcal{M}^{(2)}(q) &= \frac{1}{\hbar}\tilde I_r^{(2)}(q) + \frac{i}{2\hbar^2} \int \frac{\hat{\text{d}}^4\ell}{(2\pi)^2}\ \hat\delta(2\bar p_1\cdot\ell)\hat\delta\left(2\bar p_2\cdot(q-\ell)\right))\tilde I_r^{(1)}(\ell)\tilde I_r^{(1)}(q-\ell)\,,
\end{align}
where we have defined barred momenta as
\begin{equation}
    \bar p_1 = p_1 - \frac{1}{2}q,\qquad \bar p_2 = p_2 + \frac{1}{2}q,
\end{equation}
and $\tilde I_r(q)$ is given by
\begin{equation}
    I_r(J) = \int \hat{\text{d}}^4 q\ \hat{\delta}(2\bar p_1\cdot q)\hat\delta(2\bar p_2\cdot q)e^{\frac{i}{\hbar}q\cdot b} \tilde{I}_r(q).
\end{equation}
The 1-loop amplitude receives two contributions from the radial action: the first is a genuinely classical term $\tilde I_r^{(2)}$, while the second is a convolution of the tree-level radial action in loop-momentum space, resulting from expanding the exponential to second order. The latter scales super-classically in the soft expansion, and we can identify it as the familiar iteration term from potential scattering in non-relativistic quantum mechanics. At 1-loop, subtracting this term to isolate the classical contribution is trivial. The reason is that it is purely contained in the box and cross-box diagrams (Figure \ref{fig:diagrams} (a) and (b)), which don't contribute to the $\oh(q^{-1})$ classical order. As a result, the problem of computing the classical radial action at 1-loop is reduced to extracting the $\oh(q^{-1})$ piece of the diagrams contributing to the scalar triangles (Figure \ref{fig:diagrams} (c) and (d)).

From the classical radial action $I_r(J)$, different observables can be extracted in a straightforward way. In particular, the deflection angle is given by
\begin{equation}
    \chi(J,E) = -2\pi\left( \frac{1}{2} + \frac{\partial I_r}{\partial J} \right)\,.
\end{equation}
In this work, we are interested in analyzing the precession of orbits in a classical bound system. This, as shown in \cite{Kalin:2019rwq,Kalin:2019inp}, is related to the unbound case by direct analytic continuation of the radial action to negative energies. As a result, the periapsis advance during one orbital period can be obtained from the scattering angle as
\begin{equation}
    \label{eq:b2b}
    \Delta\Phi(J,E) = \chi(J,E) + \chi(-J,E)= -2\pi \left[\left.\frac{\partial I_r}{\partial J}\right|_{J,E}+\left.\frac{\partial I_r}{\partial J}\right|_{-J,E}\right]\,,\qquad E<0\,. 
\end{equation}
As we noted above, the classical expansion correlates powers in $G$ with powers of angular momentum $J$ (or equivalently $q$ in Fourier space). Thus, power counting makes it manifest that the only contributions to the precession of the orbit come from odd loop orders. At even orders, the deflection angle only has odd powers of the angular momentum, leading to no-precession.

Using these results, we can formulate a criterion to determine whether 2PM corrections will result in the precession of orbits (equivalently, whether the classical LRL symmetry is preserved at 2PM order). Since the classical 1-loop radial action $\tilde I_r^{(2)}(q)$ is homogeneous in the transfer momentum $q$ (and therefore, the angular momentum $J$), the only way to make $\Delta\Phi=0$ is to demand that $\tilde I_r^{(2)}(q)=0$. In other words, the $\oh(q^{-1})$ contribution to the 1-loop scattering amplitude has to vanish completely in order to preserve the LRL symmetry at the classical level. We investigate this condition in more detail in Section \ref{sec:classical}.

\subsection{Quantum potential and bound-state degeneracy}
\label{subsec:reviewquantum}
In this section, we discuss a quantum-mechanical consequence of LRL symmetry, namely bound-state degeneracy which can be extracted from a quantum potential. We also outline how we calculate quantum corrections to the potential via matching to a non-relativistic EFT.

It is well-known that the LRL symmetry of certain classical systems persists at the quantum-mechanical level; the hydrogen atom being a classic example. Here, the hidden $SO(4)$ symmetry is responsible for the energy eigenstates being degenerate in the orbital quantum number $l$, depending only on the principal number $n$ (\ref{eq:H0energies}). Further, in \cite{Caron-Huot:2018ape} such bound-state degeneracy was used to characterize the presence of quantum LRL symmetry in planar $\mathcal{N}=4$ supersymmetric Yang-Mills, where it is lifted to dual-conformal invariance in the full relativistic theory. Taking inspiration from this calculation, we will use bound-state degeneracy as a diagnostic for the presence of LRL symmetry at the first quantum order.

In order to characterize when quantum LRL symmetry is preserved, we begin by computing quantum-mechanical corrections to the scattering amplitude \eqref{eq:firstamp} by implementing the on-shell techniques we reviewed in Section \ref{subsec:reviewclassical}. Indeed, since the $\hbar$ expansion is in direct correspondence with the soft expansion of the scattering amplitude, the first quantum correction will be given by the next sub-leading contribution at small $q$. From the amplitude, we can then determine the corrections to the quantum potential $V(\mathbf{x},\mathbf{p})$ by matching between our relativistic quantum field theory to a non-relativistic EFT, following the methodology presented in \cite{Cheung:2018wkq}. This can be done to all-orders-in-velocity, which allows us to work directly in the post-Minkowskian expansion for the potential. 

Before we present the details of this matching calculation, it is valuable to understand what the different corrections to the potential are and whether they interact with one another. Note that we will work in isotropic gauge, which removes the dependence on the radial momentum $p_r\sim \frac{\mathbf{x}\cdot\mathbf{p}}{r}$ from the Hamiltonian via a canonical transformation. We can then characterize classical i.e.\@ $\mathcal{O}(\hbar^0)$ and quantum i.e.\@ $\mathcal{O}(\hbar)$ corrections to the potential by the following expansion\footnote{Note that $\mathbf{p}$ and $\mathbf{x}$ have a non-trivial commutation relation and therefore this expression is well-defined only up to a choice of operator ordering. As explained in Appendix $\ref{app:quantum energy levels}$, this subtlety does not alter the conclusion that the quantum potential must vanish to preserve the LRL degeneracy.} 
\begin{equation}\label{eq:PN expansion potential}
\begin{aligned}
    H(\mathbf{x},\mathbf{p}) = H_0(\mathbf{x},\mathbf{p}) + \sum_{m=0}^\infty \sum_{k=0}^\infty \left(\mu c^2 v^{(c)}_{k,m}+\frac{\hbar c}{r} v^{(q)}_{k,m}\right) \left(\frac{\alpha}{\mu c^2 r}\right)^m\left(\frac{\mathbf{p}^2}{\mu^2 c^2}\right)^k\,,
\end{aligned}
\end{equation}
where the unperturbed Hamiltonian (whose energy levels are degenerate) is denoted by
\begin{align}
    H_0(\mathbf{x},\mathbf{p}) = \frac{\mathbf{p}^2}{2\mu} - \frac{\alpha}{r}\,.
\end{align}
Here, $\mu := \frac{m_1 m_2}{m_1+m_2}$ is the reduced mass of the system and $\alpha$ is a dimensionful constant of units $[\alpha] = M L^3 T^{-2}$ (e.g.\@ $G\, m_1 m_2$ in the case of gravitational interactions). The terms $v^{(c)}_{k,m}$ encode the classical contributions at $m$-th PM and $k$-th PN order, while $v^{(q)}_{k,m}$ correspond to the first quantum corrections.

Our goal is to understand the conditions on the correction coefficients $v_{k,m}{(c,q)}$ such that the perturbative corrections to the energy levels of $H(\mathbf{x},\mathbf{p})$ are independent of the quantum number $l$. It turns out this requires that each correction to the Hamiltonian vanishes independently. The proof of this statement is twofold. First, we show that different orders in the Born series do not mix. Second, we show that the $l$-functional dependencies of the energy shifts due to each $v_{k,m}^{(c,q)}$ form a linearly independent set, i.e.\@ they cannot conspire to cancel unless each contribution vanishes individually. The proof that (quantum) bound-state degeneracy requires 
\begin{align}
    v_{k,m}^{(q)}=0\qquad \forall\quad m\ge 2\,, k\ge 0\,,
\end{align}
is presented in Appendix \ref{app:quantum energy levels}. In particular, for our 2PM potential to yield degenerate energy levels, we demand that $v_{2,k}=0$ for all $k\ge 0$ i.e.\@ at all-orders-in-velocity.

To establish conventions and highlight some subtle points about the definition of the effective potential, we will now briefly review the non-relativistic EFT matching definition of the PM potential introduced in \cite{Cheung:2018wkq}. We start by defining the following non-relativistic EFT in momentum space, involving two distinct massive scalars $\Phi_1$ and $\Phi_2$ interacting via a COM frame potential $V(\mathbf{p},\mathbf{p}')$
\begin{equation}
\begin{aligned}
    \label{eq:SEFT}
    S_{\text{EFT}} &:= \int \text{d}t \;\hat{\text{d}}^3\mathbf{p} \sum_{i=1,2}\Phi_i^\dagger(-\mathbf{p})\left(i\partial_t - \sqrt{\mathbf{p}^2 + m_i^2}\right)\Phi_i(\mathbf{p}) \\
   &\hspace{10mm}- \int \text{d}t \;\hat{\text{d}}^3\mathbf{p}\;\hat{\text{d}}^3\mathbf{p}' \;V(\mathbf{p}, \mathbf{p'}) \Phi_1^\dagger(\mathbf{p'}) \Phi_1(\mathbf{p}) \Phi_2^\dagger(-\mathbf{p'}) \Phi_2(-\mathbf{p}).
\end{aligned}
\end{equation}
The familiar position space potential is related to the EFT Wilson coefficient appearing in (\ref{eq:SEFT}) by a simple Fourier transform in the ``momentum transfer'' $\mathbf{q}:=\mathbf{p}'-\mathbf{p}$, that is
\begin{equation}
    V(\mathbf{x},\mathbf{p}) := \int \hat{\text{d}}^{D-1}\mathbf{q} \; e^{i\mathbf{q} \cdot\mathbf{x}} \; V(\mathbf{p},\mathbf{p}+\mathbf{q}).
\end{equation}
The potential non-relativistic EFT is defined to reproduce the physical observables of the low-velocity limit of the original fully relativistic QFT, where we have integrated out the \textit{hard} $\mathcal{O}(m)$ and \textit{soft} $\mathcal{O}(m|\mathbf{v}|)$ scales, retaining only degrees-of-freedom with momenta of order the \textit{ultra-soft} scale $\mathcal{O}(m|\mathbf{v}|^2)$; further details of the non-relativistic region expansion of the loop integrals are given in Appendix \ref{app:integrals}. The Feynman rules for this simple (bosonic) EFT are:
\begin{equation}\label{eq:EFT 2PM}
\begin{aligned}
        \begin{tikzpicture}[baseline={([yshift=-0.5ex]a.center)}]
            \begin{feynman}
                \vertex (a);
                \vertex [right=3cm of a] (b);
                \diagram* {
                    (a) -- [fermion, edge label={$(p^0, \mathbf{p})$}] (b)
                };
            \end{feynman}
        \end{tikzpicture}
    &= 
    \frac{i}{p^0 - \sqrt{\mathbf{p}^2 + m^2} + i0}, 
    \\[1em] 
    \vcenter{\hbox{
        \begin{tikzpicture}
            \begin{feynman}
                \vertex (c);
                \vertex [above left=1.2cm of c] (i1) {$\mathbf{p}$};
                \vertex [below left=1.2cm of c] (i2) {$-\mathbf{p}$};
                \vertex [above right=1.2cm of c] (f1) {$\mathbf{p}'$};
                \vertex [below right=1.2cm of c] (f2) {$-\mathbf{p}'$};
                
                \diagram* {
                    (i1) -- [fermion] (c),
                    (i2) -- [fermion] (c),
                    (c) -- [fermion] (f1),
                    (c) -- [fermion] (f2)
                };
                \filldraw[black] (c) circle (2pt); 
            \end{feynman}
        \end{tikzpicture}
    }}
    &= 
    -iV(\mathbf{p}, \mathbf{p}').
\end{aligned}
\end{equation}
There are additional complications associated with matter particle spin transitions in the presence of massless fermion mediators and other more general interactions; we defer details of this generalization of the EFT to Section \ref{sec:quantum}.  It is conventional to use the non-relativistic normalization of asymptotic states for the EFT scattering amplitude, the matching condition with the UV or full theory is therefore
\begin{equation}
    \label{eq:EFTmatching}
    M_{\text{EFT}} \overset{!}{=} \frac{\mathcal{M}_{\text{Full}}}{4E_1 E_2},
\end{equation}
which is imposed order-by-order in three separate expansions: the usual perturbative expansion in the coupling $\alpha$; the semi-classical expansion in large angular momentum $J\gg \hbar$; and (at least formally) the non-relativistic expansion $|\mathbf{p}|\ll m$. The full theory amplitude $\mathcal{M}_{\text{Full}}$ is calculated in the COM frame in the explicit kinematics
\begin{equation}
    p_1 = (E_1,\mathbf{p}),\hspace{5mm} p_2 = (E_2,-\mathbf{p}), \hspace{5mm} p_1' = (E_1,\mathbf{p}'),\hspace{5mm} p'_2 = (E_2,-\mathbf{p}'),
\end{equation}
where $E_i := \sqrt{\mathbf{p}^2+m_i^2}\,$ and we also define $E=E_1+E_2$. Up to $\mathcal{O}(\alpha^2)$ the EFT amplitude is given as the sum of tree and 1-loop bubble contributions:
\begin{equation}
    \begin{aligned}
        M_{\text{EFT}} &= 
    \vcenter{\hbox{
        \begin{tikzpicture}
            \begin{feynman}
                \vertex (c);
                \vertex [above left=1.2cm of c] (i1) {$\mathbf{p}$};
                \vertex [below left=1.2cm of c] (i2) {$-\mathbf{p}$};
                \vertex [above right=1.2cm of c] (f1) {$\mathbf{p}'$};
                \vertex [below right=1.2cm of c] (f2) {$-\mathbf{p}'$};
                
                \diagram* {
                    (i1) -- [fermion] (c),
                    (i2) -- [fermion] (c),
                    (c) -- [fermion] (f1),
                    (c) -- [fermion] (f2)
                };
                \filldraw[black] (c) circle (2pt);
            \end{feynman}
        \end{tikzpicture}
    }}
    \quad + \quad
    \vcenter{\hbox{
        \begin{tikzpicture}
            \begin{feynman}
                \vertex (v1);
                \vertex [right=2.5cm of v1] (v2); 
                \vertex [above left=1.2cm of v1] (i1) {$\mathbf{p}$};
                \vertex [below left=1.2cm of v1] (i2) {$-\mathbf{p}$};
                \vertex [above right=1.2cm of v2] (f1) {$\mathbf{p}'$};
                \vertex [below right=1.2cm of v2] (f2) {$-\mathbf{p}'$};
                
                \diagram* {
                    (i1) -- [fermion] (v1),
                    (i2) -- [fermion] (v1),
                    (v1) -- [fermion, bend left=60, edge label={$(E_1+\ell^0, \mathbf{p}+\boldsymbol{\ell})$}] (v2),
                    (v1) -- [fermion, bend right=60, edge label'={$(E_2-\ell^0, -\mathbf{p}-\boldsymbol{\ell})$}] (v2),
                    (v2) -- [fermion] (f1),
                    (v2) -- [fermion] (f2)
                };
                \filldraw[black] (v1) circle (2pt);
                \filldraw[black] (v2) circle (2pt);
                \node at (5,0) {+ ...};
            \end{feynman}
        \end{tikzpicture}
    }}
    \end{aligned}
    \label{eq:EFTtreeloop}
\end{equation}
Matching powers of the coupling, at $\mathcal{O}(\alpha)$ and $\mathcal{O}(\alpha^2)$ the amplitude in the EFT is given by
\begin{align}
    M_{\text{EFT}}^{(0)}(\mathbf{p},\mathbf{p}') &= -V^{(0)}(\mathbf{p},\mathbf{p}') \\
    \label{eq:MEFT1}
    M_{\text{EFT}}^{(1)}(\mathbf{p},\mathbf{p}') &= -V^{(1)}(\mathbf{p},\mathbf{p}')-\int \hat{\text{d}}^{D-1}\boldsymbol{\ell}\; V^{(0)}(\mathbf{p},\mathbf{p}+\boldsymbol{\ell}) \Delta(\mathbf{p}+\boldsymbol{\ell}) V^{(0)}(\mathbf{p}+\boldsymbol{\ell},\mathbf{p}'),
\end{align}
the second term in the 1-loop expression is variously described as the \textit{iteration} or \textit{Born subtraction} contribution \cite{Cheung:2018wkq,Cristofoli:2019neg}. To calculate the leading quantum correction to $V^{(1)}$ we need to expand the effective propagator in (\ref{eq:MEFT1}) to NNLO in the soft limit
\begin{equation}\label{eq:EFT prop expanded}
    \begin{aligned}
        \Delta(\mathbf{p}+\boldsymbol{\ell}) &= \frac{1}{E-\sqrt{E_1^2+\boldsymbol{\ell}^2+2\mathbf{p}\cdot\boldsymbol{\ell}}-\sqrt{E_2^2+\boldsymbol{\ell}^2+2\mathbf{p}\cdot\boldsymbol{\ell}}+i0}\\
        &=-\frac{\xi E}{\mathbf{p}\cdot\boldsymbol{\ell}{-}i0} + \frac{\xi E\boldsymbol{\ell}^2}{2(\mathbf{p}\cdot\boldsymbol{\ell}{-}i0)^2} - \frac{1-3\xi}{2\xi E} +  \frac{1-4\xi+\xi^2}{4(\xi E)^3}(\mathbf{p}\cdot\boldsymbol{\ell}) - \frac{\xi E}{4}\frac{(\boldsymbol{\ell}^2)^2}{(\mathbf{p}\cdot\boldsymbol{\ell}{-}i0)^3} + \oh(|\boldsymbol{\ell}|^{2}),
    \end{aligned}
\end{equation}
where we have defined $\xi := \frac{E_1 E_2}{E^2}$. 

The matching condition (\ref{eq:EFTmatching}) holds on-shell, which in the COM frame means it holds on the support of the energy conservation condition $\mathbf{p}^2=\mathbf{p'}^2$ or equivalently $\mathbf{p}\cdot\mathbf{q} = -\frac{1}{2}\mathbf{q}^2$. We can use this condition to remove any dependence on the invariant $\mathbf{p}\cdot\mathbf{q}$, up to higher-order terms in the soft expansion. The choice to define the potential as a function of the invariants $\mathbf{p}^2$ and $\mathbf{q}^2$ coincides with the definition of isotropic gauge described above and greatly simplifies the resulting expression. Without loss of generality, the soft expansion of the (non-analytic in $|\mathbf{q}|$ part of the) isotropic gauge potential takes the general form
\begin{equation}\label{eq:EFT potential PM expansion}
\begin{aligned}
    V(\mathbf{p},\mathbf{p}+\mathbf{q}) = \frac{\alpha c_1(\mathbf{p}^2)}{|\mathbf{q}|^2}+ \frac{\alpha^2 c_2(\mathbf{p}^2)}{|\mathbf{q}|} + \hbar\alpha^2 c_3(\mathbf{p}^2)\log(|\mathbf{q}|)+\ldots,
\end{aligned}
\end{equation}
where each $c_i$ encodes an infinite number of terms in the PN expansion. In position space the potential takes the equivalent form
\begin{align}
    V(\mathbf{x},\mathbf{p}) = \frac{\alpha c_1(\mathbf{p}^2)}{4 \pi r} + \frac{\alpha^2 c_2(\mathbf{p}^2)}{2\pi^2 r^2} -\frac{\hbar \alpha^2 c_3(\mathbf{p}^2)}{4\pi r^3}+\cdots,
\end{align}
where we have used the standard Fourier transforms
\begin{equation}
    \label{eq:Fourier}
    \int \hat{\text{d}}^{3}\mathbf{q} \; e^{i\mathbf{q} \cdot\mathbf{x}} \; \frac{1}{\mathbf{q}^2} = \frac{1}{4\pi r}, \hspace{5mm} \int \hat{\text{d}}^{3}\mathbf{q} \; e^{i\mathbf{q} \cdot\mathbf{x}} \; \frac{1}{|\mathbf{q}|} = \frac{1}{2\pi^2 r^2}, \hspace{5mm} \int \hat{\text{d}}^{3}\mathbf{q} \; e^{i\mathbf{q} \cdot\mathbf{x}} \; \log(|\mathbf{q}|) = -\frac{1}{4\pi r^3}.
\end{equation}
As discussed above, and argued in more detail in Appendix \ref{app:quantum energy levels}, the condition to preserve the LRL degeneracy at the quantum level requires $c_3(\mathbf{p}^2)=0$. The possible realization of this condition in an explicit model is analyzed in detail in Section \ref{sec:quantum}.

\subsection{EFT of long-range forces}
\label{sec:eft_ops}

To study the preservation of LRL symmetry in a generic theory, we need to first enumerate all possible interactions that contribute classically and quantum-mechanically to the scattering between two massive scalar particles $\Phi_{1},\,\Phi_{2}$. In this section, we present the on-shell matrix elements for all the interaction vertices that we consider.

\begin{figure}
\centering

\begin{tikzpicture}[baseline=(current bounding box.center)]
    \draw[thick] (0,1.2) -- (2,1.2);
    \draw[thick] (0,0) -- (2,0);
    \draw[scalarnoarrow,thick] (0.5,1.2) -- (0.5,0);
    \draw[scalarnoarrow,thick] (1.5,1.2) -- (1.5,0);
    \node at (1,-0.5) {(a)};
\end{tikzpicture}
\qquad
\begin{tikzpicture}[baseline=(current bounding box.center)]
    \draw[thick] (0,1.2) -- (2,1.2);
    \draw[thick] (0,0) -- (2,0);
    \draw[scalarnoarrow,thick] (0.5,1.2) -- (1.5,0);
    \draw[scalarnoarrow,thick] (1.5,1.2) -- (0.5,0);
    \node at (1,-0.5) {(b)};
\end{tikzpicture}
\qquad
\begin{tikzpicture}[baseline=(current bounding box.center)]
    \draw[thick] (0,1.2) -- (2,1.2);
    \draw[thick] (0,0) -- (2,0);
    \coordinate (vertex) at (1,0.6);
    \draw[scalarnoarrow,thick] (0.5,1.2) -- (vertex);
    \draw[scalarnoarrow,thick] (1.5,1.2) -- (vertex);
    \draw[scalarnoarrow,thick] (vertex) -- (1,0);
    \node at (1,-0.5) {(c)};
\end{tikzpicture}
\qquad
\begin{tikzpicture}[baseline=(current bounding box.center)]
    \draw[thick] (0,1.2) -- (2,1.2);
    \draw[thick] (0,0) -- (2,0);
    \draw[scalarnoarrow,thick] (0.5,1.2) -- (1,0);
    \draw[scalarnoarrow,thick] (1.5,1.2) -- (1,0);
    \node at (1,-0.5) {(d)};
\end{tikzpicture}

\par\medskip

\begin{tikzpicture}[baseline=(current bounding box.center)]
    \draw[thick] (0,1.2) -- (2,1.2);
    \draw[thick] (0,0) -- (2,0);
    \draw[scalarnoarrow,thick] (1,0.6) circle[radius=0.3];
    \draw[scalarnoarrow,thick] (1,1.2) -- (1,0.9);
    \draw[scalarnoarrow,thick] (1,0.3) -- (1,0);
    \node at (1,-0.5) {(e)};
\end{tikzpicture}
\qquad\qquad
\begin{tikzpicture}[baseline=(current bounding box.center)]
    \draw[thick] (0,1.2) -- (2,1.2);
    \draw[thick] (0,0) -- (2,0);
    \draw[scalarnoarrow,thick] (1,0.84) circle[radius=0.36];
    \draw[scalarnoarrow,thick] (1,0.48) -- (1,0);

    \node at (1,-0.5) {(f)};
\end{tikzpicture}
\qquad\qquad
\begin{tikzpicture}[baseline=(current bounding box.center)]
    \draw[thick] (0,1.2) -- (2,1.2);
    \draw[thick] (0,0) -- (2,0);
    \draw[scalarnoarrow,thick] (1,0.6) circle[radius=0.6];

    \node at (1,-0.5) {(g)};
\end{tikzpicture}

\caption{Topologies contributing to the classical (upper row) and quantum (lower row) amplitude.}
\label{fig:diagrams}
\end{figure}
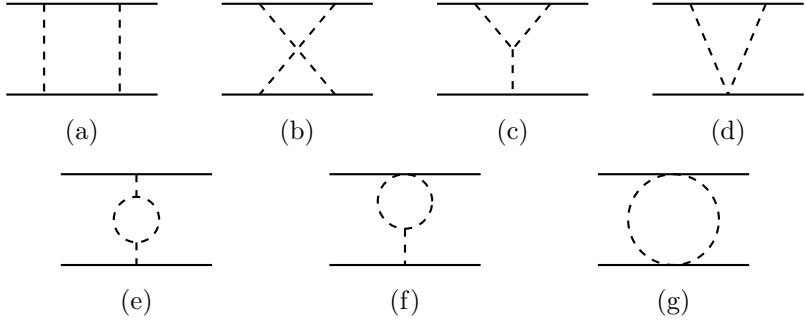

As previously mentioned, the classical contribution to the amplitude corresponds to a $\oh(q^{-1})$ scaling in the soft expansion, while the quantum piece behaves as $\oh(q^0)$. This restricts the possible types of exchange diagrams that contribute to the ones presented in Figure \ref{fig:diagrams}, where the classical piece only includes the upper row of graphs.
By counting powers of the soft momentum in each diagram, it is clear that three-point vertices involving two massive scalars and one massless field can have at most two powers of soft momentum, while vertices involving three massless bosons must scale at most as $\oh(q^3)$. Meanwhile, four-point interactions involving two massive scalars and two massless mediators can only contain a single power of soft momenta. 

Finally, some naively relevant interactions have been excluded on physical grounds. The first such is a Yukawa coupling between a massless scalar and two massless fermions. From soft scaling alone, one can show that diagrams such as the Y-diagram (c) in Figure \ref{fig:diagrams} actually contribute at classical order $\oh(q^{-1})$. We exclude such an interaction as we do not expect fermion exchanges to play a role in classical physics. In a similar vein, we exclude couplings such as $\phi^p$ ($p\geq3$) and interactions involving non-Abelian gauge fields. The former would give rise to forces scaling as $r^{2p-5}$, while the latter would run into the problem of confinement at large distances, despite behaving like a long-range potential $\log(r)/r$ in a perturbative expansion at small coupling. It would be interesting to see if there is a more rigorous argument to exclude both of these types of contributions.

Taking these considerations into account, we enumerate the list of \textit{relevant} interactions in Table \ref{tab:3pt massive class} for the interactions between the massive matter particles and the massless mediators and Table \ref{tab:3pt massless class} for self-interactions between the massless mediators. In each case we provide the (lowest-multiplicity) on-shell matrix element needed to define the associated coupling constants/Wilson coefficients. The required 1-loop integrand can be either directly constructed using unitarity cuts, or first matched to an off-shell effective action and calculated using Feynman rules, possibly including ghost fields in a covariant gauge.
\begin{table}[t]
    \centering
    \renewcommand{\arraystretch}{0.7}
    \begin{tabular}{| Sc | Sc |}
        \hline
         External states $(\boldsymbol{s_1},\boldsymbol{s_2},...,\boldsymbol{s_n})$ & On-shell matrix element $i\mathcal{M}_n(p_1,...,p_n)$\\
        \hline
        $(\mathbf{0},\bar{\mathbf{0}},0)$ & $-ig$\\
        \hline
        $(\mathbf{0},\bar{\mathbf{0}},1)$ & $-iQ(\varepsilon_3\cdot p_1 - \varepsilon_3\cdot p_2)$\\
        \hline
        $(\mathbf{0},\bar{\mathbf{0}},2)$ & $i\kappa\, (\varepsilon_3\cdot p_1)(\varepsilon_3\cdot p_2)$\\
        \hline
        $\left( \displaystyle \mathbf{0},\bar{\mathbf{\frac{1}{2}}},\frac{1}{2} \right)$ & $i g_Y\, (\bar{u}_2 v_3)$ \\
        \hline
        $\left( \displaystyle \bar{\mathbf{0}},\mathbf{\frac{1}{2}},\frac{3}{2} \right)$ & $-\displaystyle\frac{i\kappa}{4} (\varepsilon_3\cdot p_1 - \varepsilon_3\cdot p_2)(\bar u_3 v_2)$ \\
        \hline
        $(\mathbf{0},\bar{\mathbf{1}},0)$ & $i w (\varepsilon_2\cdot q_3)$ \\
        \hline
        $(\mathbf{0},\bar{\mathbf{1}},1)$ & $-i\tau \left[ (\varepsilon_2 \cdot q_3) (\varepsilon_3\cdot p_2) - (\varepsilon_2\cdot\varepsilon_3)(p_2 \cdot q_3) \right]$ \\
        \hline
        $(\mathbf{0},\bar{\mathbf{0}},0,0)$ & $-ih$ \\
        \hline
        $\left( \displaystyle\mathbf{0},\bar{\mathbf{0}},\frac{1}{2},\frac{1}{2} \right)$ & $-ih_Y (\bar{u}_3 v_4)$ \\
        \hline
    \end{tabular}
    \caption{Interactions between the massive particles and the massless mediators of a given spin $s_i$. In the left column, a \textbf{bold} spin indicates that the particle is massive, and a bar indicates Hermitian conjugation. Momenta $p_i$ are defined in the all-outgoing convention; the graviton polarization tensor and gravitino polarization vector-spinor have been written in a ``double-copy'' form $\varepsilon^\mu \varepsilon^\nu$ and $\varepsilon^\mu v(p)$ respectively, where we have implicitly imposed the traceless ($\varepsilon^2 =0$) and gamma-trace ($\slashed{\varepsilon}(p)v(p) =0$) conditions. The gravitational coupling is defined as $\kappa = \sqrt{32\pi G}$.}
    \label{tab:3pt massive class}
\end{table}
\begin{table}[t]
    \centering
    \begin{tabular}{| Sc | Sc |}
        \hline
       External states $(s_1,s_2,s_3)$ & On-shell matrix element $i\mathcal{M}_3(p_1,p_2,p_3)$ \\
        \hline
       $(0,0,2)$ & $\displaystyle i\kappa (\varepsilon_3\cdot q_1)(\varepsilon_3\cdot q_2)$\\
        \hline
        $(0^+,1,1)$& $-i\lambda \left[ (q_2 \cdot q_3) (\varepsilon_2\cdot \varepsilon_3) - (\varepsilon_3\cdot q_2)(\varepsilon_2\cdot q_3) \right]$\\
        \hline
        $(0^-,1,1)$& 
        $-i\tilde{\lambda} \epsilon^{\mu\nu\rho\sigma} q_{2\rho} q_{3\sigma}\varepsilon_{2\mu}\varepsilon_{3\nu}$  \\
        \hline
        $(1,1,2)$ & $i\kappa(\varepsilon_3\cdot q_1)\left[ (\varepsilon_1\cdot\varepsilon_2)(\varepsilon_3\cdot q_1) + (\varepsilon_1\cdot\varepsilon_3)(\varepsilon_2\cdot q_3) + (\varepsilon_2\cdot\varepsilon_3)(\varepsilon_1\cdot q_2) \right]$ \\
        \hline
         $(2,2,2)$ & $-i\kappa\left[ (\varepsilon_1\cdot\varepsilon_2)(\varepsilon_3\cdot q_1) + (\varepsilon_1\cdot\varepsilon_3)(\varepsilon_2\cdot q_3) + (\varepsilon_2\cdot\varepsilon_3)(\varepsilon_1\cdot q_2) \right]^2$ \\
        \hline
        $\left(\displaystyle\frac{1}{2},\frac{1}{2},0\right)$ & $-i \zeta (\bar{u}_1 v_2)$ \\
        \hline
        $\left(\displaystyle\frac{1}{2},\frac{1}{2},1\right)$ & $-i Q_f \bar{u}_1 \slashed{\varepsilon}_3 v_2 -i\alpha_f \bar{u}_1[\slashed{q}_3,\slashed{\varepsilon}_3]v_2$ \\
        \hline
        $\left(\displaystyle\frac{1}{2},\frac{1}{2},2\right)$ & $\displaystyle\frac{i\kappa}{8} (\varepsilon_3\cdot q_2 - \varepsilon_3\cdot q_1) (\bar{u}_1\slashed{\varepsilon}_3 v_2)$ \\
        \hline
        $\left(\displaystyle\frac{3}{2},\frac{3}{2},2\right)$ & $\displaystyle\frac{i\kappa}{4}(\bar u_1 \slashed{\varepsilon}_3 v_2)\left[(\varepsilon_1\cdot\varepsilon_3)(\varepsilon_2\cdot q_3) + (\varepsilon_2\cdot\varepsilon_3)(\varepsilon_1\cdot q_2) - (\varepsilon_1\cdot\varepsilon_2)(\varepsilon_3\cdot q_1) \right]$ \\
        \hline
        $\left(\displaystyle\frac{3}{2},\frac{1}{2},1\right)$ & $\displaystyle \frac{i\kappa g_Y}{2Q}(\bar u_1\slashed{\varepsilon_3} v_2)(\varepsilon_1\cdot q_2)$ \\
        \hline
    \end{tabular}
    \caption{Three-point self-interactions between massless mediators. For interactions with external states $(0,1,1)$ we have distinguished parity even (scalar)/parity odd (pseudo-scalar) interactions as $+$ and $-$ respectively.}
    \label{tab:3pt massless class}
\end{table}

\section{Classical LRL Symmetry: No Orbital Precession}
\label{sec:classical}

In this section we derive the necessary and sufficient \textit{classical} conditions for the existence of LRL symmetry in a two-body system. In Section \ref{sec:pre_cond}, we discuss the general condition (see \eqref{eq:tricut}) that must be satisfied in a theory with interactions of the types discussed in Section \ref{sec:eft_ops}. In Section \ref{sec:BPS_KK}, we discuss the no-precession condition for a BPS-$\overline{\text{BPS}}$ bound state. Finally in Section \ref{sec:Nge4}, we show that any $\mathcal{N}\ge4$ supergravity has classical LRL symmetry, as opposed to $\mathcal{N}=8$ alone \cite{Caron-Huot:2018ape}.

\subsection{Conditions for no-precession}
\label{sec:pre_cond}

In the context of the general EFT of long-range interactions described in Section \ref{sec:eft_ops}, as might be expected, at $\mathcal{O}(\hbar^0)$ only bosonic mediators contribute to PM scattering observables. To simplify the presentation, we begin with an analysis of a model containing as mediators a spin-2 graviton, in addition to a single spin-1 massless vector and spin-0 massless scalar with generic charges, self-couplings and couplings to the massive spin-0 matter particles. This theory is a good starting point for our discussion: simple enough to make the LRL conditions tractable, yet complex enough to illustrate how the more general calculation will proceed. We are interested in the classical radial action for a $2\to2$ scattering process of the two massive scalar particles. 

As reviewed in Section \ref{subsec:reviewclassical}, we calculate the 1-loop classical radial action from the corresponding 1-loop amplitude, constructed by sewing the $t$-channel as in \eqref{eq:comptoncut}. The classical contribution is then proportional to the $\oh(q^{-1})$ term in the soft expansion, all together we find the explicit result
\begin{equation}
\begin{aligned}
    \tilde{I}_r^{(2)}(q)=&\frac{1}{512\,m_1\sqrt{-q^2} }\Bigg(2 g_1^2 \left(\kappa^2 m_2^2-8 h_2\right)+8 m_1^2 Q_1^2 \left(\kappa^2 m_2^2+4 g_2 \lambda +16 Q_2^2\right)-3 \kappa^4 m_2^2 m_1^4\\
    &\hspace{20mm}-32 m_1 m_2 Q_1 Q_2 \sigma \left(2g_1 \lambda +3 \kappa^2 m_1^2\right)-\kappa^2 m_2^2 \sigma^2 \left(2g_1^2+3m_1^2 \left(8 Q_1^2- 5 \kappa^2 m_1^2\right)\right)\Bigg)\\
    &\hspace{20mm}+ (1\leftrightarrow2),
\end{aligned}
\end{equation}
where the definitions of the coupling constants appearing in this expression are given in Tables \ref{tab:3pt massive class} and \ref{tab:3pt massless class}. As argued in Section \ref{subsec:reviewclassical}, using the boundary-to-bound continuation formula (\ref{eq:b2b}), the absence of precession requires the 2PM radial action itself to vanish at all orders in the velocity (PN) expansion. Therefore we impose that the coefficient of each power of $\sigma$ is zero; this gives the following no-precession conditions
\begin{equation}
\begin{aligned}
\label{eq:classicalLRLcond}
    3 \kappa^4 m_1^4 m_2^2 +2g_1^2 \left(8h_2-\kappa^2 m_2^2\right)-8 m_1^2Q_1^2 \left(\kappa^2 m_2^2+ 16 Q_2^2+4 g_2 \lambda\right) + (1\leftrightarrow 2) = 0,\\
    Q_1Q_2\left[ m_1\left( 3\kappa^2 m_2^2 + 2g_2 \lambda \right)+(1\leftrightarrow 2) \right] =0,\\
    m_2^3\left[ 2g_1^2+3 m_1^2 \left(8Q_1^2-5\kappa^2  m_1^2\right) \right]+(1\leftrightarrow 2)=0.
\end{aligned}
\end{equation}
It is instructive to consider some simple limits of these equations where we restrict the spectrum of allowed mediators. Setting $g_i=0$ decouples the scalar field; the resulting model is just the familiar \textit{Einstein-Maxwell} theory; the constraints for no-precession in this case reduce to
\begin{equation}
\begin{aligned}
    3 \kappa^4 m_1^4 m_2^2-8 m_1^2 Q_1^2 \left(\kappa^2 m_2^2+ 16 Q_2^2\right) + (1\leftrightarrow 2) = 0,\\
    Q_1Q_2\left( m_1 +m_2  \right) =0,\\
    m_2\left(8Q_1^2-5\kappa^2  m_1^2\right) + (1\leftrightarrow 2)=0.
\end{aligned}
\end{equation}
Since $m_i>0$, the second condition requires $Q_1 Q_2=0$; without loss of generality we will choose $Q_2=0$. Solving the remaining equations gives a unique non-trivial solution
\begin{equation}
\label{eq:solEM}
  m_2=2m_1, \hspace{10mm} |Q_1| = \sqrt{\frac{15}{8}}\,\kappa m_1.
\end{equation}
In other words, a bound system of a Schwarzschild and a Reissner-Nordstr\"{o}m black hole with this particular electric charge and mass ratio has zero precession through 2PM. Conversely, for generic mass ratios these constraints have no solutions and so we conclude: \textit{for generic mass ratios it is impossible to solve the no-precession conditions without the presence of a massless scalar mediator}. 

Similarly we can consider the case where we set $Q_i=0$, decoupling the photon giving a kind of \textit{scalar-tensor} theory. In this case the constraints reduce to 
\begin{equation}
\begin{aligned}
    \label{eq:LRLcondnovector}
    2g_2^2 \left(\kappa^2m_1^2- 8h_1\right)-3\kappa^4 m_1^2 m_2^4 + (1\leftrightarrow 2) = 0,\\
    m_2^3\left(2g_1^2-15 \kappa^4 m_1^4\right)+ (1\leftrightarrow 2)=0.
\end{aligned}
\end{equation}
There are many solutions to these conditions, including solutions that are valid for all mass ratios.\footnote{This conclusion corrects an error in the analysis of \cite{Davis:2023zqv}, where it is asserted that the LRL symmetry is necessarily broken at 2PN in scalar-tensor models. The calculation in that paper is missing the cubic interaction between the graviton and the massless scalar required by the equivalence principle. We are grateful to Scott Melville for communicating with us and confirming this.}

An equivalent way to state the conditions (\ref{eq:classicalLRLcond}) is to isolate the $\mathcal{O}(\hbar^0)$ part of the 1-loop amplitude in the usual scalar integral basis (see Appendix \ref{app:integrals} for the explicit expressions)
\begin{equation}
    \mathcal{M}_4^{\text{1-loop}}\biggr\vert_{\mathcal{O}(\hbar^0)} = c^{(0)}_{\tric{}} I^{(-1)}_{\tric{}} + c^{(0)}_{\tricinv{}} I^{(-1)}_{\tricinv{}}.
\end{equation}
where the superscript $(...)^{(n)}$ corresponds to the $\mathcal{O}(q^n)$ contribution in the soft expansion. Here we are using the fact that the soft expansions of the scalar box integrals $I_{\nb{}}$ and $I_{\cb{}}$ have no contribution at $\mathcal{O}(q^{-1})$. The coefficients $c_{\tric{}}$, $c_{\tricinv{}}$ can be related directly to the tree-level amplitudes through the corresponding triangle cuts \cite{Forde:2007mi}. Since the triangle integrals are equal up to a factor of $m_i$, the classical condition for no-precession can simply be stated as the requirement that the leading term in a certain linear combination of triangle coefficients of the 1-loop amplitude vanishes
\begin{equation}
\label{eq:tricut}
\boxed{m_2 c^{(0)}_{\tric{}} + m_1 c^{(0)}_{\tricinv{}}= 0\,.}
\end{equation}
A stronger (and optional) version of this constraint consists in imposing that each triangle coefficient vanishes independently: $c^{(0)}_{\tric{}}=c^{(0)}_{\tricinv{}}=0$. An even stronger version is the so-called ``no-triangle property'' present in some models such as $\mathcal{N}=8$ SUGRA \cite{Arkani-Hamed:2008owk, Caron-Huot:2018ape}; in such models the triangle coefficients are exactly zero, whereas no-precession requires only that they are zero at leading order in the soft expansion. The stronger version of the constraint however appears to be necessary if we insist on the vanishing of precession for arbitrary mass ratios; by contrast the peculiar Einstein-Maxwell solution (\ref{eq:solEM}), which fixes $m_1/m_2 = 1/2$, requires a cancellation between the two different triangle coefficients.

It is straightforward to generalize this discussion to a model with multiple flavors of photons and scalars. Consider a theory with $N_s$ flavors of scalar and $N_v$ flavors of vector particles. To extend our calculation of the radial action presented above, we can account for the multiple species by simply replacing couplings in each diagram topology as
\begin{equation}
    Q_i\rightarrow Q_{i,I},\hspace{7mm} g_i \rightarrow g_{i,a},\hspace{7mm} h_i\rightarrow h_{i,ab},\hspace{7mm} \lambda \rightarrow \lambda_{aIJ},
\end{equation}
where $i=1,2$ corresponds to the body-$i$, $a=1,...,N_s$ is a (real) scalar species label and $I,J=1,...,N_v$ is a vector species label. Under such replacements, our conditions \eqref{eq:classicalLRLcond} become
\begin{equation}
    \begin{aligned}
        3\kappa^4 m_1^4 m_2^2 + 2\sum_{a,b=1}^{N_s}g_{1,a}g_{1,b}\left( 8\,h_{2,ab} - \delta_{ab}\,\kappa^2 m_2^2\right)\hspace{50mm}\\
        -8m_1^2 \sum_{J,K=1}^{N_v}Q_{1,J}Q_{1,K}\left[ \kappa^2 m_2^2\delta_{JK}+16Q_{2,J}Q_{2,K} + 4\sum_{a=1}^{N_s}g_{2,a}\lambda_{aJK} \right] + (1\leftrightarrow 2) &= 0,\\
         \sum_{J,K=1}^{N_v}Q_{1,J}Q_{2,K}\left[m_1\left(3\kappa^2 m_1^2 \delta_{JK} + \sum_{a=1}^{N_s}2g_{1,a}\lambda_{aJK}  \right) + (1\leftrightarrow 2)\right] &=0,\\
        m_2^3\left[\sum_{a=1}^{N_s}2g_{1,a}^2 - 3m_1^2\left( 5\kappa^2 m_1^2-\sum_{J=1}^{N_v}8Q_{1,J}^2 \right)\right]+ (1\leftrightarrow 2) &= 0.
    \end{aligned}
\end{equation}
These are the versions of the no-precession condition \eqref{eq:tricut}, that we will use in Section \ref{sec:Nge4} when discussing $4\leq \mathcal{N}\le 8$ supergravity multiplets, whose $R$-symmetry necessitates the existence of multiple flavors of scalars and photons.

\subsection{BPS-\texorpdfstring{$\overline{\text{BPS}}$}{antiBPS} bound states and Kaluza-Klein}
\label{sec:BPS_KK}
In all claimed examples of systems with a hidden LRL symmetry \cite{Caron-Huot:2014gia, Caron-Huot:2018ape, deNeeling:2023egt}, the particles forming the two-body bound state saturate a \textit{BPS} or \textit{extremality} bound. There is no \textit{a priori} reason that this should be required in a more general model. Still, in order to make contact with the results in this section, we consider the fate of our constraints \eqref{eq:classicalLRLcond} on the support of extremality conditions.

It is important to define what we mean by ``extremal'' here. If the static force between a pair of identical particles vanishes, we consider the two-body system to have saturated an extremality or BPS bound. All spherically symmetric black hole solutions to two-derivative equations of motion at zero Hawking temperature have this property \cite{Heidenreich:2020upe}; as do (BPS) Kaluza-Klein modes in dimensional reductions of (super-)Yang-Mills and (super-)gravity \cite{Scherk:1979aj}. 

To obtain the no-force condition, we calculate the tree-level potential between two identical states ($m_1=m_2:= m$, $Q_1=Q_2:= Q$, $g_1=g_2:= g$) and demand that it vanishes in the static limit $\sigma\to1$. Explicitly we find
\begin{equation}
    V_{\text{tree}}(\mathbf{x},\mathbf{p}) \overset{\sigma \rightarrow 1}{=} -\frac{ \left(\kappa^2 m^4-8Q^2 m^2+2g^2\right)}{16\pi r }\,.
\end{equation}
Thus, the tree-level extremality condition reads $g_i^2=m_i^2\left(4Q_i^2-\frac{1}{2}\kappa^2m_i^2\right)$\footnote{One can also write the tree-level no-force condition for multiple flavors of photons and scalars as
\begin{equation}
    \sum_{a=1}^{N_s}g_a^2 = m^2\left(\sum_{J=1}^{N_v}8Q_J^2-\frac{1}{2}\kappa^2m^2\right)\,.
\end{equation}
}. Imposing the same condition at 1-loop fixes the scalar ``seagull'' coupling $h$ defined in Table \ref{tab:3pt massive class}.

We can now consider a two-body system formed by a pair of extremal particles $(m_i,Q_i,g_i)$ satisfying the above condition but with electric charges of \textit{opposite sign} i.e.\@ a non-canceling static force. This is a ``$\text{BPS}$-$\overline{\text{BPS}}$'' bound state. Together with the general conditions for the absence of precession (\ref{eq:classicalLRLcond}) this gives
\begin{equation}
    Q_i = \pm \frac{1}{\sqrt{2}}\kappa\, m_i\,,\qquad h_i = \frac{1}{2}\kappa^2m_i^2\,,\qquad \lambda =\sqrt{\frac{3}{2}}\,\kappa\,,
\end{equation}
where we have additionally assumed that the corrections to the radial action vanish for all possible values of the mass ratio $m_1/m_2$. 

Interestingly, these two conditions, i.e.\@ extremality and no-precession fix the value of the bulk $\varphi F_{\mu\nu}^2$ coupling $\lambda$ completely. This specific value coincides with the dilaton coupling to the graviphoton field resulting from the Kaluza-Klein (KK) reduction of 5D Einstein gravity \cite{Ortin:2004ms,Cornish:1996de}\footnote{Since we want to compare with a scattering amplitude calculation, here we give the KK action with the less common ``particle physics'' normalization for the scalar and gauge fields with engineering dimensions $[\varphi]=[A_\mu]=1$ and standard kinetic term pre-factors in the $(+,-,-,-)$ metric signature. }
\begin{equation}
    S_{\text{KK}}[g_{\mu\nu},A_\mu, \varphi] = \int \text{d}^4 x \sqrt{-g}\left[\frac{2}{\kappa^2} R +\frac{1}{2}\left(\nabla_\mu \varphi\right)^2 - \frac{1}{4} e^{-\kappa\sqrt{\frac{3}{2}}\varphi} F_{\mu\nu}^2\right],
\end{equation}
consistent with the claims of \cite{deNeeling:2023egt}. This might suggest that one actually needs the full exponential form of the gauge field's kinetic term for the symmetry to hold at all orders in the PM expansion, which would be interesting to explore in higher loop calculations.

\subsection{No-precession for \texorpdfstring{$\mathcal{N}\geq$4}{N>=4} SUGRA}
\label{sec:Nge4}
In this section, we consider another set of solutions to the no-precession condition \eqref{eq:tricut} -- this time in the context of the two-body $\frac 12$-BPS classical potential in  $4\leq \mathcal{N}\le 8$ supergravity. The most symmetric case is that of $\mathcal{N}=8$ supergravity which was discussed in detail in \cite{Caron-Huot:2018ape}. Indeed as we show in this section, $\mathcal{N}=8$ is not the unique choice that allows us to satisfy \eqref{eq:tricut} and in fact any $4\leq \mathcal{N}\le 8$ satisfies it.

The amplitude we are interested in is an $4\leq \mathcal{N}\le 8$, 4-point 1-loop amplitude with $\frac 12$-BPS scalar external states. This can be constructed via a dimensional reduction of the massless superamplitude \cite{Parra-Martinez:2020dzs} from $D=6$ to $D=4$\footnote{We focus here on a particular set of BPS external states i.e.\@ the case of single-angle misalignment, since this is the one readily accessible via dimensional reduction.}. To do this, one chooses the massless polarization vectors to lie in the hyperplane perpendicular to the four physical dimensions in which the momenta of the external particles lie. We forgo a detailed discussion of this procedure here; this can be found in \cite{Parra-Martinez:2020dzs}. One notable aspect of this procedure is that it does not change the topology of a given scalar integral. While some topologies vanish under dimensional reduction\footnote{In essence, this is due to momentum conservation in the extra dimensions which forces mass (or Kaluza-Klein number) conservation at the vertices. If there is no way to conserve mass in a given topology, then the diagram gives zero upon dimensional reduction.}, others are unchanged i.e.\@ boxes remain boxes, triangles remain triangles, etc.

Equipped with dimensional reduction, we have reduced the problem to the construction of massless superamplitudes in $4\leq \mathcal{N}\le 8$ supergravity. These are given by the Bern-Carrasco-Johansson (BCJ) double copy of $\mathcal{N}=4$ supersymmetric Yang-Mills (SYM) with $ \mathcal{N}\le 4$ SYM. At 4-point 1-loop
\begin{align}
\label{eq:Nge4}
     \mathcal{M}_4^{4+\mathcal{N}\; \text{SUGRA}} =-\left(\frac{\kappa}{2}\right)^4 \int\frac{\text{d}^4 \ell}{(2\pi)^4} \left[\frac{n^{\mathcal{N}=4}_{1234} n^{\mathcal{N}}_{1234}}{d_{1234}}+\frac{n^{\mathcal{N}=4}_{1423} n^{\mathcal{N}}_{1423}}{d_{1423}}+\frac{n^{\mathcal{N}=4}_{1342} n^{\mathcal{N}}_{1342}}{d_{1342}}\right]\,,
\end{align}
where we have introduced scalar box integral propagators $d_{abcd}$,
\begin{align}
     d_{abcd} = \ell^2 (\ell-k_b)^2 (\ell - k_b-k_c)^2 (\ell+k_a)^2\,,
\end{align}
the color-kinematic numerators of generic $\mathcal{N}\leq 4$ SYM, denoted $n_{ijkl}^{\mathcal{N}}$, and those in $\mathcal{N}=4$ SYM:
\begin{align}
\label{eq:N4num}
    n^{\mathcal{N}=4}_{1234}= s_{12}s_{14} \mathcal{A}_4^\text{tree}(k_1, k_2, k_3,k_4)\,.
\end{align}
Note that in \eqref{eq:Nge4}, we have dropped all non-box numerators. This is because $\mathcal{N}=4$ SYM only has non-zero box numerators and so the other cubic topologies in the second copy, including triangles, may be neglected. This does not immediately imply the vanishing of the triangle coefficient due to non-trivial dependence of the box numerators on loop momentum.

The explicit form of the BCJ numerators for $\mathcal{N}<4$ are given in equations (4.7), (4.10) and (4.11) of \cite{Carrasco:2012ca}. Unlike their $\mathcal{N}=4$ SYM counterpart \eqref{eq:N4num}, they are $\ell$-dependent. Therefore, in order to check whether or not the triangle coefficient vanishes, one must perform integration-by-parts reduction. Explicit calculation yields non-zero box and bubble coefficients for $4\le \mathcal{N}<8$ supergravity, but a vanishing triangle coefficient. 

In order to compare this with the condition \eqref{eq:tricut}, we need to further Kaluza-Klein reduce these massless amplitudes. At classical order, this amounts to setting:
\begin{align}
     I_{\bub{}}=0\,, && I_{1234}=I_{\nb{}}\,, && I_{1423}=0\,, && I_{1342} = I_{\cb{}}\,.
\end{align}
Importantly, as we noted before, this dimensional reduction does not change the topology of the Feynman integral i.e.\@ there remain no-triangle coefficients even after we reduce. Interestingly, the classical limit of the $4\leq \mathcal{N}\le 8$ SUGRA amplitude after KK reduction matches the $\mathcal{N}=8$ SUGRA amplitude exactly, i.e. it is given by
\begin{align}
    \mathcal{M}_4^{\mathcal{N}\ge 4 \text{ SUGRA}} = -\kappa^4 m_1^4 m_2^4 (\sigma+1)^4 (I_{\nb{}}+I_{\cb{}})\,.
\end{align}
Thus we conclude that the no-triangle condition (and hence the no-precession condition) is satisfied by $\frac12$-BPS classical potentials in $4\leq \mathcal{N}\le 8$ supergravity.

\section{Quantum LRL Symmetry: Hydrogen-like Degeneracy}
\label{sec:quantum}

As we have seen, the condition that LRL symmetry is preserved by classical relativistic corrections is only mildly constraining and is satisfied in a large class of models with vector and scalar exchanges. In this section, the central new result of the present paper, we extend this analysis to the leading quantum corrections; finding that the conditions are much more non-trivial to satisfy. As discussed in Section \ref{subsec:reviewquantum}, preserving the quantum LRL symmetry is expected to imply that the quantum two-body bound states exhibit the same $l$-independent degeneracy as the hydrogen atom, and as argued in detail in Appendix \ref{app:quantum energy levels} this in turn is equivalent to the vanishing of the leading quantum correction to the potential. 

In Section \ref{subsec:boson qu potential}, using an extension of the same on-shell techniques as for the classical case, we explain how to efficiently extract the 2PM quantum potential from relativistic scattering amplitudes in a completely generic low-energy EFT of quantum gravity and compute the corresponding conditions for the preservation of the quantum LRL symmetry. In passing from classical to quantum PM effects, we encounter qualitatively new features, the most important of which is the presence of long-range forces mediated by massless fermions and internal spin-transitions. As we will see, massless spin-$\frac{3}{2}$ or gravitino exchanges play a uniquely important role in the ultra-relativistic limit of quantum scattering, and in Section \ref{subsec:gravitino} we are led to the remarkable observation that the cancellation of the graviton contribution to the quantum potential necessarily requires exactly 6 Majorana gravitinos i.e.\@ $\mathcal{N}=6$ supergravity. Finally in Section \ref{subsec:N8sugra} we revisit the conjecture of \cite{Caron-Huot:2018ape}, showing that bound states in $\mathcal{N}=8$ supergravity do not preserve the quantum degeneracy and explicitly calculate the leading fine-structure correction to the energy levels. Note that there is one important caveat with this approach: we only consider quantum potentials with sensible expansions in the low-velocity limit. We discuss this further in Section \ref{sec:discussion}.

\subsection{Quantum post-Minkowskian potential}
\label{subsec:boson qu potential}
The calculation of the leading quantum correction to the effective potential follows from the EFT approach reviewed in Section \ref{sec:review}, extended to NNLO in the soft expansion (\ref{eq:softexpansion}). The 2PM $\mathcal{O}(\alpha^2)$ scattering amplitude in the non-relativistic EFT is composed of a tree and a 1-loop iteration diagram \eqref{eq:EFTtreeloop}. Expanding both the potential coefficients $c_i$ in \eqref{eq:EFT potential PM expansion} and the effective two-body propagator \eqref{eq:EFT prop expanded}, at order $\oh(|\mathbf{q}|^0)$ we have
\begin{align}
    M_\text{EFT}(\mathbf{p},\mathbf{p}+\mathbf{q})\biggr\vert_{\mathcal{O}(\alpha^2\hbar)}&=-\alpha^2 c_3(\mathbf{p}^2) \log(|\mathbf{q}|) \nonumber\\
    &\hspace{5mm}+\frac{\alpha^2 E\xi [c_1(\mathbf{p}^2)]^2}{4}\int \hat{\text{d}}^{D-1} \boldsymbol{\ell}\; \frac{\boldsymbol{\ell}^2}{(\boldsymbol{\ell}-\mathbf{q})^2[\mathbf{p}\cdot \boldsymbol{\ell}-i0]^3} \nonumber\\
    &\hspace{5mm}+ \alpha^2 \biggr(-\frac{\left(\xi^2-4\xi +1\right)}{4E^3 \xi^3}[c_1(\mathbf{p}^2)]^2+\frac{3\xi-1}{E \xi}c_1(\mathbf{p}^2)c_1'(\mathbf{p}^2)\nonumber\\
    &\hspace{15mm}+2E \xi c_1(\mathbf{p}^2)c_1''(\mathbf{p}^2) \biggr)\int \hat{\text{d}}^{D-1} \boldsymbol{\ell}\; \frac{\mathbf{p}\cdot\boldsymbol{\ell}}{\boldsymbol{\ell}^2(\boldsymbol{\ell}-\mathbf{q})^2}\,.
\end{align}
The remaining integrals are finite in $D=4$ and can be straightforwardly evaluated:
\begin{equation}
    \int \hat{\text{d}}^{3} \boldsymbol{\ell}\; \frac{\boldsymbol{\ell}^2}{(\boldsymbol{\ell}-\mathbf{q})^2[\mathbf{p}\cdot \boldsymbol{\ell}-i0]^3} = 0,\hspace{5mm} \int \hat{\text{d}}^{3} \boldsymbol{\ell}\; \frac{\mathbf{p}\cdot\boldsymbol{\ell}}{\boldsymbol{\ell}^2(\boldsymbol{\ell}-\mathbf{q})^2} = \mathcal{O}(|\mathbf{q}|)\,.
\end{equation}
The only assumption being made in this calculation is that the tree-level potential is defined in isotropic gauge, otherwise our calculation is model-independent. We conclude then that the \textit{quantum iteration contribution is zero for bosonic mediators without spin transitions}. The quantum correction to the 2PM potential is simply
\begin{equation}
    \label{eq:Vquantum}
    V_{\text{bosonic}}(\mathbf{x},\mathbf{p})\biggr\vert_{\mathcal{O}(\alpha^2 \hbar)} = - \frac{1}{4E_1 E_2}  \int \hat{\text{d}}^{3}\mathbf{q} \; e^{i\mathbf{q} \cdot\mathbf{x}} \; \left(\mathcal{M}_4^{\text{1-loop}}(\mathbf{p},\mathbf{p}+\mathbf{q}) \biggr\vert_{\log(|\mathbf{q}|)}\right).
\end{equation}
A corollary of the above claim is that the quantum part of the amplitude must be purely real and regular in the static $\sigma\rightarrow 1$ limit. Below we confirm this prediction in a few representative examples. As we will see in the following subsection this statement is slightly modified in the case of fermionic mediators or bosonic mediators with internal spin-transitions. In each case there is a purely imaginary iteration contribution that must be subtracted, the prescription (\ref{eq:Vquantum}) is simply modified by taking the real part of $\mathcal{M}_4^{\text{1-loop}}$ on the right-hand-side.

\subsubsection*{Example 1: 2PL quantum potential in scalar QED }

A simple non-gravitational example of quantum potential matching is the two-photon exchange potential in scalar QED at second \textit{post-Lorentzian} (2PL) order or $\mathcal{O}(\alpha_{\text{em}}^2 \hbar)$\footnote{The electromagnetic fine structure constant $\alpha_{\text{em}}$ and (dimensionless) electric charges $e_i$ are defined here such that the tree-level Coulomb potential takes the standard form $V_{\text{Coulomb}}(r) = -\frac{\alpha_{\text{em}} e_1 e_2}{r}$.} 

\begin{equation}
    \label{eq:quantum2PLQED}
    V_{\text{QED}}(\mathbf{x},\mathbf{p})\biggr\vert_{\mathcal{O}(\alpha_{\text{em}}^2 \hbar)} = -\frac{\alpha_{\text{em}}^2 e_1^2 e_2^2}{2 \pi r^3}\frac{\sqrt{\sigma
   ^2-1} \left(\sigma ^2+1\right)+2 \sigma 
   \left(\sigma ^2-2\right) \text{arccosh}(\sigma
   )}{E_1(\mathbf{p}) E_2(\mathbf{p})
   \left(\sigma ^2-1\right)^{3/2}}.
\end{equation}
As explained in Section \ref{sec:review}, all potentials in this paper are given in isotropic gauge. This expression was calculated in scalar QED, but it is reasonable to expect that it agrees with the spin-independent part of the potential in spinor QED. The static limit ($\sigma\rightarrow 1$) of this result was first calculated many decades ago \cite{Feinberg:1988yw}, and then more recently using modern unitarity methods in \cite{Holstein:2008sw,Bjerrum-Bohr:2013bxa}; from our result \eqref{eq:quantum2PLQED} we find 
\begin{equation}
    V_{\text{QED}}(\mathbf{x},\mathbf{p})\biggr\vert_{\mathcal{O}(\alpha_{\text{em}}^2\hbar)}\overset{\sigma \rightarrow 1}{=} -\frac{7}{6 \pi }\frac{\alpha_{\text{em}} ^2 e_1^2 e_2^2}{ 
   m_1 m_2 r^3},
\end{equation}
in complete agreement with the expected expression. 

\subsubsection*{Example 2: 2PM quantum potential in general relativity }

We have also calculated the $\mathcal{O}(G^2 \hbar)$ quantum potential between a pair of minimally coupled non-spinning massive bodies in perturbatively quantized general relativity
\begin{equation}
    \label{eq:quantum2PMGR}
    V_{\text{GR}}(\mathbf{x},\mathbf{p})\biggr\vert_{\mathcal{O}(G^2\hbar)} = \frac{G^2 m_1^2 m_2^2}{30\pi  r^3 }\frac{ \sqrt{\sigma ^2-1} \left(18 \sigma ^4-49 \sigma ^2+1\right)-30 \sigma  \left(12 \sigma ^4-20 \sigma
   ^2+7\right) \text{arccosh}\left(\sigma\right)}{E_1(\mathbf{p}) E_2(\mathbf{p})\left(\sigma ^2-1\right)^{3/2}}.
\end{equation}
As previously emphasized \cite{BjerrumBohr:2002kt}, since the long-range part of the potential depends only on contributions to the scattering amplitude that are non-analytic as $q^2\rightarrow 0$, it is therefore insensitive to the details of the UV physics and is a universal low-energy prediction of quantum gravity. In the static limit $\sigma\rightarrow 1$ this reduces to the famous result \cite{BjerrumBohr:2002kt,Bjerrum-Bohr:2013bxa}
\begin{equation}
    V_{\text{GR}}(\mathbf{x},\mathbf{p})\biggr\vert_{\mathcal{O}(G^2\hbar)} \overset{\sigma \rightarrow 1}{=} -\frac{41}{10 \pi} \frac{G^2 m_1 m_2 }{r^3}.
\end{equation}
As far as we are aware, the all-orders-in-velocity (PL/PM) results (\ref{eq:quantum2PLQED}), (\ref{eq:quantum2PMGR}) for the potentials in 
QED and GR have not previously appeared in the literature. The corresponding quantum part of the 1-loop eikonal gravitational amplitude was calculated in \cite{Bjerrum-Bohr:2021din}, and agrees with the above expression (\ref{eq:quantum2PMGR}).

\subsection{Long-range fermion exchange forces}
\label{sec:fermion}
At subleading order in the $\hbar$ expansion, there are additional contributions to the quantum potential from Feynman diagrams where the massive internal lines become higher spin particles. In particular, allowing the massive scalars to transition into spin-1/2 and spin-1 particles generates a tree-level amplitude that is already of order $\oh(\hbar)$, and thus will produce an iteration term that scales as $\log(|\mathbf{q}|)$ and mixes with the 2PM potential. However, as will be demonstrated, this iteration term is purely imaginary and can thus be straightforwardly separated when matching to the relativistic amplitude.

To account for these contributions, we need to enhance the non-relativistic EFT to include massive scalars, fermions and vectors (all with the same mass), as well as an interaction potential between them. Since the calculations for all of these are very similar, we will present the details for the case of gravitino exchange, computing the iteration term to the bound-state potential at 1-loop and comparing it to the full amplitude result in Appendix \ref{app:gravitino details}.

\subsubsection*{Example: gravitino exchange}

As an illustrative example (of importance for the discussion in the following section), we will calculate the quantum potential generated by exchanging a pair of spin-3/2 gravitinos. As discussed, this is constructed by sewing a pair of Compton amplitudes with external massless spin-3/2 particles. The details of the construction of these Compton amplitudes, and the sewing are given in Appendix \ref{app:gravitino details}; the result is the following 1-loop quantum amplitude
\begin{align}\label{eq:gravitino amplitude}
    \mathcal{M}_4^{\text{1-loop}}\biggr\vert_{\text{gravitino}} &= \frac{\kappa ^4 m_1^2 m_2^2 \left(2
   \sqrt{\sigma ^2-1} \left(47 \sigma ^2-14\right)+30 \sigma  \left(4 \sigma ^2-3\right) \text{arccosh}\left(\sigma\right)\right)}{7680 \pi ^2 \sqrt{\sigma ^2-1}}\log\left(-q^2\right) \nonumber\\
   &\hspace{10mm} -\frac{i\kappa ^4 m_1^2 m_2^2 \left(4 \sigma ^3-3 \sigma +1\right)}{512 \pi \sqrt{\sigma ^2-1}}\log\left(-q^2\right),
\end{align}
where for simplicity we are only displaying terms that are non-analytic in $q^2$. This expression cannot be immediately matched to an effective potential since the quantum amplitude has a non-zero imaginary part. For consistency this must match the iteration of the tree-level fermion potential. The non-relativistic EFT Lagrangian \eqref{eq:SEFT} defined
in Section \ref{subsec:reviewquantum} must be enlarged to contain all possible interaction potentials between pairs of massive scalars and spin-1/2 fermions
\begin{align}
    L_{\text{EFT}} &= \sum_{a=1}^2\int \hat{\text{d}}^3\mathbf{p}\left[\Phi_a^\dagger({-}\mathbf{p})\left(i\partial_t {-} \sqrt{\mathbf{p}^2+m_a^2}\right)\Phi_a(\mathbf{p}) + \sum_{\pm}\Psi_{\pm,a}^\dagger({-}\mathbf{p}) \left(i\partial_t {-} \sqrt{\mathbf{p}^2+m_a^2}\right) \Psi_{\pm,a}(\mathbf{p}) \right] \nonumber\\
    &\hspace{5mm}  - \int \hat{\text{d}}^3\mathbf{p}'\;\hat{\text{d}}^3\mathbf{p} \; V_{\Phi\Phi\rightarrow\Phi\Phi}\left(\mathbf{p},\mathbf{p}'\right) \Phi^\dagger_1(\mathbf{p}')\Phi^\dagger_2(-\mathbf{p}')\Phi_1(\mathbf{p})\Phi_2(-\mathbf{p}) \nonumber\\
    &\hspace{5mm} - \sum_{\pm}\int \hat{\text{d}}^3\mathbf{p}'\;\hat{\text{d}}^3\mathbf{p} \; V_{\Phi\Phi\rightarrow\Psi_\pm \Psi_\mp}\left(\mathbf{p},\mathbf{p}'\right) \Psi_{\pm,1}^\dagger(\mathbf{p}')\Psi_{\mp,2}^\dagger(-\mathbf{p}')\Phi_1(\mathbf{p})\Phi_2(-\mathbf{p}) \nonumber\\
    &\hspace{5mm}  - \sum_{\pm}\int \hat{\text{d}}^3\mathbf{p}'\;\hat{\text{d}}^3\mathbf{p} \; V_{\Psi_\pm \Psi_\mp \rightarrow \Phi\Phi}\left(\mathbf{p},\mathbf{p}'\right) \Psi_{\pm,1}(\mathbf{p})\Psi_{\mp,2}(-\mathbf{p})\Phi^\dagger_1(\mathbf{p}')\Phi^\dagger_2(-\mathbf{p}') \nonumber\\
    &\hspace{5mm} +\;\; ...
\end{align}
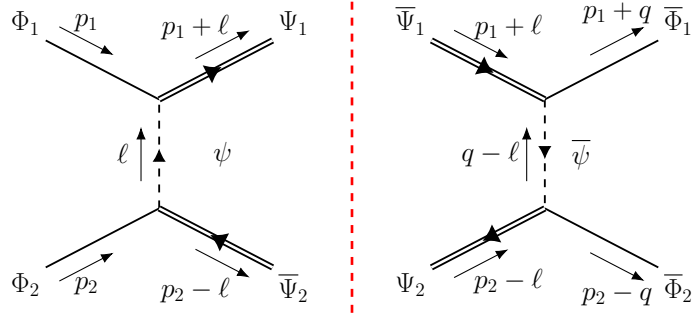
\begin{figure}
\centering
\begin{tikzpicture}[
    scale=0.6,
    transform shape,
    >=Latex,
    line width=0.7pt,
    midarr/.style={
        postaction={decorate},
        decoration={
            markings,
            mark=at position 0.55 with {
                \arrow{Latex[length=6pt,width=8.1pt]}
            }
        }
    },
    midarrMassless/.style={
        postaction={decorate},
        decoration={
            markings,
            mark=at position 0.55 with {
                \arrow{Latex[length=5pt,width=5.1pt]}
            }
        }
    }
]
\begin{scope}
    \coordinate (A-vt) at (0,1.2);
    \coordinate (A-vb) at (0,-1.2);

    \coordinate (A-phi1) at (-2.5,2.5);
    \coordinate (A-phi2) at (-2.5,-2.5);
    \coordinate (A-psi1) at (2.5,2.5);
    \coordinate (A-psibar2) at (2.5,-2.5);
    \node[above left]  at (A-phi1)    {\LARGE$\Phi_1$};
    \node[below left]  at (A-phi2)    {\LARGE$\Phi_2$};
    \node[above right] at (A-psi1)    {\LARGE$\Psi_1$};
    \node[below right] at (A-psibar2) {\LARGE$\overline{\Psi}_2$};
    \node at (1.4,0) {\LARGE$\psi$};
    \draw (A-phi1) -- (A-vt);
    \draw (A-phi2) -- (A-vb);

    \draw[dashed,midarrMassless] (A-vb) -- (A-vt);

    \draw[double,double distance=1pt,midarr]
        (A-vt) -- (A-psi1);

    \draw[double,double distance=1pt,midarr]
        (A-psibar2) -- (A-vb);
    \draw[->,thin] (-2.2,2.8) -- (-1.0,2.176);
    \node at (-1.6,2.9) {\LARGE$p_1$};

    \draw[->,thin] (-2.2,-2.8) -- (-1.0,-2.176);
    \node at (-1.6,-2.9) {\LARGE$p_2$};

    \draw[->,thin] (0.8,2.1) -- (2.0,2.724);
    \node at (0.8,2.8) {\LARGE$p_1+\ell$};

    \draw[->,thin] (0.8,-2.1) -- (2.0,-2.724);
    \node at (0.8,-3) {\LARGE$p_2-\ell$};

    \draw[->,thin] (-0.4,-0.6) -- (-0.4,0.6);
    \node at (-0.8,0) {\LARGE$\ell$};

\end{scope}

\draw[red, dashed, line width=1pt]
    (4.25,-3.5) -- (4.25,3.5);
\begin{scope}[shift={(8.5,0)}]
    \coordinate (B-vt) at (0,1.2);
    \coordinate (B-vb) at (0,-1.2);

    \coordinate (B-psi1) at (-2.5,2.5);
    \coordinate (B-psibar2) at (-2.5,-2.5);
    \coordinate (B-phi1) at (2.5,2.5);
    \coordinate (B-phi2) at (2.5,-2.5);
    \node[above left]  at (B-psi1)    {\LARGE$\overline{\Psi}_1$};
    \node[below left]  at (B-psibar2) {\LARGE$\Psi_2$};
    \node[above right] at (B-phi1)    {\LARGE$\overline{\Phi}_1$};
    \node[below right] at (B-phi2)    {\LARGE$\overline{\Phi}_2$};
    \node at (0.8,0) {\LARGE$\overline{\psi}$};
    \draw (B-vt) -- (B-phi1);
    \draw (B-vb) -- (B-phi2);
    \draw[dashed,midarrMassless] (B-vt) -- (B-vb);
    \draw[double,double distance=1pt,midarr]
        (B-psi1) -- (B-vt);

    \draw[double,double distance=1pt,midarr]
        (B-vb) -- (B-psibar2);
    \draw[<-,thin] (2.2,2.8) -- (1.0,2.176);
    \node at (1.6,3.1) {\LARGE$p_1+q$};

    \draw[<-,thin] (2.2,-2.8) -- (1.0,-2.176);
    \node at (1.6,-3.1) {\LARGE$p_2-q$};

    \draw[<-,thin] (-0.8,2.1) -- (-2.0,2.724);
    \node at (-0.8,2.8) {\LARGE$p_1+\ell$};

    \draw[<-,thin] (-0.8,-2.1) -- (-2.0,-2.724);
    \node at (-0.8,-2.8) {\LARGE$p_2-\ell$};

    \draw[->,thin] (-0.4,-0.6) -- (-0.4,0.6);
    \node at (-1.2,0) {\LARGE$q-\ell$};

\end{scope}

\end{tikzpicture}
\caption{Example of a gravitino exchange diagram at tree-level and its conjugate, where the massive scalar $\Phi$ transitions into a spin-1/2 fermion $\Psi$ by emitting/absorbing a massless spin-3/2 gravitino $\psi$. Gluing both together results in the iteration term of the quantum potential.}
\label{fig:gravitino tree-level}
\end{figure}
Here the operators $\Phi_i$ and $\Psi_{\pm,i}$ annihilate (non-relativistically normalized) one-particle states. In particular, $\Psi_{+,i}$ annihilates a fermion of mass $m_i$, and $\Psi_{-,i}$ annihilates the corresponding anti-fermion. These species-changing ``potentials'' are calculated by matching the tree-level amplitudes such as the one shown in the left of Figure \ref{fig:gravitino tree-level}, where the massive scalars exchange a single gravitino, transitioning into massive spin-1/2 fermion states. To keep consistent with the choices we made for the scalar case, we work in the \emph{quasi-isotropic} gauge \cite{Vines:2018gqi}, which is defined such that the coefficient of each spin structure (such as $\mathbf{r}\cdot\boldsymbol{\sigma}$ or $\mathbf{L}\cdot\boldsymbol{\sigma}$) is independent of the radial momentum $p_r$.

In the full relativistic theory, the relevant three-point interaction is given in Table \ref{tab:3pt massive class}; at leading-order in the soft expansion, the (long-range part of the) relativistic tree-level amplitudes are given by
\begin{align}
\label{eq:tree-level gravitino}
    \mathcal{M}^{\text{tree}}_4(\bar\Phi_{-p_1},\Psi_{p_1+q},\bar\Phi_{-p_2},\bar\Psi_{p_2-q}) &= \frac{\kappa^2}{8}m_1 m_2(1-2\sigma)\frac{\overline{u}_1(p_1)\slashed{q}v_2(p_2)}{q^2}+\mathcal{O}\left(q^0\right)\,,\nonumber\\
    \mathcal{M}^{\text{tree}}_4(\bar\Psi_{-p_1},\Phi_{p_1+q},\bar\Psi_{-p_2},\Phi_{p_2-q}) &= \frac{\kappa^2}{8}m_1 m_2(1-2\sigma)\frac{\overline{v}_2(p_2)\slashed{q} u_1(p_1)}{q^2}+\mathcal{O}\left(q^0\right)\,.
\end{align}
Reversing the direction of fermion flow simply corresponds to replacing $u\leftrightarrow v$. 

Importantly, the tree-level amplitudes \eqref{eq:tree-level gravitino} scale as $\oh(q^{-1})$, in contrast to the $\oh(q^{-2})$ scaling of the Coulomb/Newton potential, in the case of boson exchange. As a result, when iterating the tree-level potential over a 1-loop diagram in the non-relativistic EFT, the leading-order contribution contains the $\log\left(\mathbf{q}^2\right)$ iteration contribution
\begin{equation}
    \int \hat{\text{d}}^{3-2\epsilon}\boldsymbol{\ell}\ \underbrace{V_{\text{fermion}}^{(0)}(\boldsymbol{\ell})}_{\mathcal{O}(\mathbf{q}^{-1})}\times \underbrace{\Delta(\boldsymbol{\ell})}_{\mathcal{O}(\mathbf{q}^{-1})}\times \underbrace{V_{\text{fermion}}^{(0)}(\mathbf{q}-\boldsymbol{\ell})}_{\mathcal{O}(\mathbf{q}^{-1})} \;\;\;\sim \;\;\; \log\left(\mathbf{q}^2\right).
\end{equation}
The tree-level potentials can be straightforwardly obtained from (\ref{eq:tree-level gravitino}) by matching. When taking the non-relativistic limit, it is convenient to use the Dirac basis for the gamma matrices
\begin{equation}
    \gamma^0 = \begin{pmatrix}
        1 && 0 \\
        0 && -1
    \end{pmatrix}\,, \hspace{10mm} \gamma^i = 
    \begin{pmatrix}
        0 && \sigma^i \\
        -\sigma^i && 0
    \end{pmatrix}\,,
\end{equation}
where the corresponding explicit forms of the fermion wavefunctions are given by
\begin{equation}
    u_s(p) = \sqrt{E_{\mathbf{p}}+m}\begin{pmatrix}
        \xi_s\\
        \frac{\mathbf{p}\cdot \boldsymbol{\sigma}}{E_{\mathbf{p}}+m}\xi_s
    \end{pmatrix}\,,
    \hspace{10mm} v_s(p) = \sqrt{E_{\mathbf{p}}+m}\begin{pmatrix}
        \frac{\mathbf{p}\cdot \boldsymbol{\sigma}}{E_{\mathbf{p}}+m}\eta_s\\
        \eta_s
    \end{pmatrix}\,,
\end{equation}
and the two-component Pauli spinors are normalized as $\xi_r^{\dagger} \xi_s = \eta_r^{\dagger} \eta_s = \delta_{rs}$. In these conventions, after some simple algebra, we calculate the tree-level potentials
\begin{align}
V^{(0)}_{\Phi\Phi\rightarrow\Psi_+ \Psi_-}\left(\mathbf{p},\mathbf{p}'\right) &= \frac{\kappa^2 m_1^{3/2} m_2^{3/2}(1-2\sigma)\sqrt{\sigma+1}}{16\sqrt{2} E_1 E_2 }\frac{(\mathbf{p}'-\mathbf{p})\cdot \boldsymbol{\sigma}}{(\mathbf{p}'-\mathbf{p})^2}\,, \nonumber\\
V^{(0)}_{\Psi_- \Psi_+ \rightarrow \Phi\Phi}\left(\mathbf{p},\mathbf{p}'\right) &= \frac{\kappa^2 m_1^{3/2} m_2^{3/2}(1-2\sigma)\sqrt{\sigma+1}}{16\sqrt{2} E_1 E_2 }\frac{(\mathbf{p}'-\mathbf{p})\cdot \boldsymbol{\sigma}}{(\mathbf{p}'-\mathbf{p})^2}\,.
\end{align}
The reactions with the fermions replaced with their anti-fermions, $V_{\Phi\Phi\rightarrow\Psi_- \Psi_+}$ and $V_{\Psi_+ \Psi_- \rightarrow \Phi\Phi}$ are identical. From these expressions we calculate the iteration contribution to the potential 
\begin{align}
    &-\int \hat{\text{d}}^{3-2\epsilon}\boldsymbol{\ell}\; V^{(0)}_{\Phi\Phi\rightarrow \Psi_+\Psi_-}(\mathbf{p},\mathbf{p}+\boldsymbol{\ell})\Delta(\mathbf{p},\mathbf{p}+\boldsymbol{\ell})V^{(0)}_{\Psi_-\Psi_+\rightarrow \Phi\Phi }(\mathbf{p}+\boldsymbol{\ell},\mathbf{p}+\mathbf{q}) \nonumber\\
    &= -\frac{\kappa^4 m_1^3 m_2^3\left(1-2\sigma\right)^2\left(\sigma+1\right)}{512 E_1 E_2(E_1+E_2)}\int \hat{\text{d}}^{3-2\epsilon}\boldsymbol{\ell} \frac{\text{Tr}\left[\left(\boldsymbol{\ell}\cdot \boldsymbol{\sigma}\right)\left((\boldsymbol{\ell}-\mathbf{q})\cdot \boldsymbol{\sigma}\right)\right]}{\boldsymbol{\ell}^2 (\boldsymbol{\ell}-\mathbf{q})^2\left[\mathbf{p}\cdot \boldsymbol{\ell}-i0\right]} \nonumber\\
    &= -\frac{i\kappa ^4 m_1^2 m_2^2 \left(1-2\sigma\right)^2\left(\sigma+1\right)}{2048\pi E_1 E_2 \sqrt{\sigma ^2-1}}\log\left(\mathbf{q}^2\right)\,.
\end{align}
Subtracting this from the full-theory loop amplitude (\ref{eq:gravitino amplitude}) exactly cancels the imaginary part. Importantly, the iteration contribution is \textit{purely imaginary}, and so as expected the Born subtraction is equivalent to taking the real part of the quantum amplitude. Since we only keep the leading order of the numerators and the effective propagator in the soft expansion, the result is in one-to-one correspondence with the box cut of the relativistic amplitude. With this result, the 2PM quantum potential contribution from the exchange of a pair of Majorana gravitinos is 
\begin{equation}
    V_{\text{gravitino}}(\mathbf{x},\mathbf{p})\biggr\vert_{\mathcal{O}(G^2\hbar)} = \frac{G^2 m_1^2 m_2^2}{60 \pi r^3}\frac{ 2
   \sqrt{\sigma ^2-1} \left(47 \sigma ^2-14\right)+30 \sigma  \left(4 \sigma ^2-3\right) \text{arccosh}\left(\sigma\right)}{ E_1(\mathbf{p}) E_2(\mathbf{p})\sqrt{\sigma ^2-1}}.
\end{equation}
In a theory of $\mathcal{N}$-extended SUGRA, the contribution to the potential is given by this result multiplied by a factor of $\mathcal{N}$, in addition to contributions from gravitons and lower-spin exchange.

\subsubsection*{General spin transition structure}

\begin{figure}
\centering
\begin{tikzpicture}[
    scale=0.6,
    transform shape,
    >=Latex,
    line width=0.7pt,
    midarr/.style={
        postaction={decorate},
        decoration={markings, mark=at position 0.55 with {\arrow{Latex[length=6pt, width=8.1pt]}}}
    },
    midarrMassless/.style={
        postaction={decorate},
        decoration={markings, mark=at position 0.55 with {\arrow{Latex[length=5pt, width=5.1pt]}}}
    }
]
    \coordinate (vt) at (0, 1.2);
    \coordinate (vb) at (0, -1.2);
    
    \coordinate (phiA_start) at (-2.5, 2.5);
    \coordinate (phiB_start) at (-2.5, -2.5);
    \coordinate (psiA_end) at (2.5, 2.5);
    \coordinate (psiB_bar_end) at (2.5, -2.5);
    \node at (phiA_start) [above left] {\LARGE$\Phi_1$};
    \node at (phiB_start) [below left] {\LARGE$\Phi_2$};
    \node at (psiA_end) [above right] {\LARGE$W_1^\mu$};
    \node at (psiB_bar_end) [below right] {\LARGE$\Phi_2$};
    \node at (1.4, 0) {\LARGE$\varphi$}; 
    \draw (phiA_start) -- (vt);
    \draw (phiB_start) -- (vb);
    \draw[dashed] (vb) -- (vt);
    \draw[double] (vt) -- (psiA_end);
    \draw[] (psiB_bar_end) -- (vb);

\end{tikzpicture}
\hspace{1cm}
\begin{tikzpicture}[
    scale=0.6,
    transform shape,
    >=Latex,
    line width=0.7pt,
    midarr/.style={
        postaction={decorate},
        decoration={markings, mark=at position 0.55 with {\arrow{Latex[length=6pt, width=8.1pt]}}}
    },
    midarrMassless/.style={
        postaction={decorate},
        decoration={markings, mark=at position 0.55 with {\arrow{Latex[length=5pt, width=5.1pt]}}}
    }
]
    \coordinate (vt) at (0, 1.2);
    \coordinate (vb) at (0, -1.2);
    \coordinate (phiA_start) at (-2.5, 2.5);
    \coordinate (phiB_start) at (-2.5, -2.5);
    \coordinate (psiA_end) at (2.5, 2.5);
    \coordinate (psiB_bar_end) at (2.5, -2.5);
    \node at (phiA_start) [above left] {\LARGE$\Phi_1$};
    \node at (phiB_start) [below left] {\LARGE$\Phi_2$};
    \node at (psiA_end) [above right] {\LARGE$W_1^\mu$};
    \node at (psiB_bar_end) [below right] {\LARGE$\Phi_2$};
    \node at (1.4, 0) {\LARGE$A^\nu$};
    \draw (phiA_start) -- (vt);
    \draw (phiB_start) -- (vb);
    \draw[wiggle] (vb) -- (vt);
    \draw[double] (vt) -- (psiA_end);
    \draw[] (psiB_bar_end) -- (vb);
\end{tikzpicture}
\caption{The quantum contribution to the potential also includes processes where the massive scalar line transitions into a massive spin-1 particle.}
\label{fig:vector transition}
\end{figure}
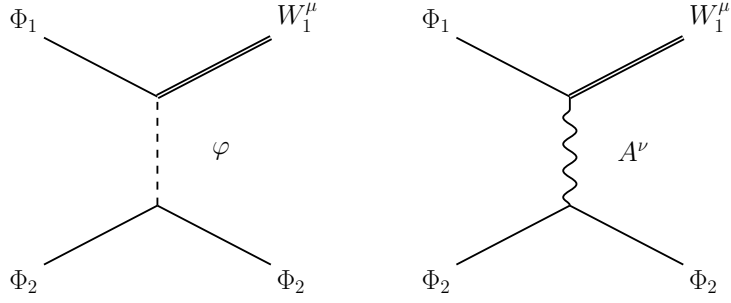

Aside from the gravitino exchange, the quantum iteration term of the potential also receives contributions associated to exchanging massless spin-1/2 mediators $\chi$, as well as processes where one of the massive scalar lines transitions into a massive vector $W^\mu$ as shown in Figure \ref{fig:vector transition}.
The computation follows exactly the same steps as in the gravitino case by using the 3-point interactions in Tables \ref{tab:3pt massive class} and \ref{tab:3pt massless class}, where in the vector case iterating over the tree-level potential involves summing over all possible massive internal spin-1 states.  Here we only present the final result accounting for all possible contributions of the massless mediators, as in the gravitino example above the Born iteration part of the potential is purely imaginary
\begin{equation}\label{eq:iteration potential}
    \begin{aligned}
        V(\mathbf{x},\mathbf{p})\biggr\vert_{\mathcal{O}(G^2\hbar)} \supset \frac{i G^2 }{8\pi^2 r^3}\frac{\left(\sigma+1\right)}{ E_1 E_2 \sqrt{\sigma ^2-1}}\Bigg[& 64|g_Y|^4 + 4m_1 m_2 |g_Y|^2(2\sigma-1)+ m_1^2 m_2^2(2\sigma -1)^2 \\ 
        &+\frac{16}{m_1 m_2}\big(g_1 \omega_2 {+} g_2\omega_1 {-} 2m_1 m_2 \sigma (Q_1\tau_2 {+} Q_2\tau_1)\big)^2\Bigg].
    \end{aligned}
\end{equation}
Subtracting this iteration contribution from the total scattering amplitude, one obtains a real 1-loop quantum potential as written in \eqref{eq:EFT potential PM expansion}.

\subsection{Ultra-relativistic expansion and graviton-gravitino cancellation}
\label{subsec:gravitino}

Finally, we have all of the ingredients ready to calculate the quantum correction to the potential in the generic long-range EFT introduced in Section \ref{sec:eft_ops} and state the necessary conditions for the preservation of the quantum LRL degeneracy. We extract the coefficient of $\log(-q^2)$ from the $t$-channel discontinuity of the relativistic amplitude, corresponding to the $\oh(\hbar)$ correction to the long-range potential between the two massive scalar bodies. The classical expansion of the amplitude organizes itself naturally in terms of the integral topologies. After gluing internal massless states in the generalized unitarity cut and reducing the integrals via integration-by-parts identities, the coefficients of the master integrals depend solely on $q^2$ and $\sigma$. Since only the box, cross-box and bubble integrals contribute to even orders in $q$, the first quantum correction to the amplitude can be written as
\begin{equation}
    \mathcal{M}^{\text{1-loop}}_4\biggr\vert_{\mathcal{O}(\hbar)} = c_{\nb{}}^{(2)} I_{\nb{}}^{(-2)} + c_{\cb{}}^{(2)}I_{\cb{}}^{(-2)} + c_{\cb{}}^{(0)}I_{\cb{}}^{(0)} + c_{\bub{}}^{(0)} I_{\bub{}}^{(0)},
\end{equation}
where the superscript $(...)^{(n)}$ corresponds to the $\mathcal{O}(q^n)$ contribution in the soft expansion and the integrals are defined by \eqref{eq:fullint}. 

After subtracting the piece of the amplitude associated with the iteration of the tree-level potential (which, as explained previously, is simply its imaginary part), we are left with the contribution from the 1-loop quantum part of the potential \eqref{eq:EFT potential PM expansion}. To preserve the LRL symmetry at 2PM order we demand that this correction vanishes to all-orders-in-velocity. From the explicit expressions for the soft expanded integrals in Appendix \ref{app:integrals}, we see that the box $I_{\nb{}}$, $I_{\cb{}}$ and bubble $I_{\bub{}}$ contributions have different functional dependence on $\sigma$ that cannot cancel. In particular the box contributions grow as $\sim \sigma^k\log(\sigma)$ in the ultra-relativistic limit $\sigma\rightarrow \infty$, whereas the bubble contributions have only power-law behavior $\sim \sigma^k$. We must therefore require the separate cancellation of the $\log(\sigma)$ contributions (equivalently the coefficients of $\text{arccosh}(\sigma)$), leading to a very simple condition on the soft expansions of the four-particle cuts
\begin{equation}
\label{eq:qu arccosh condition}
    \boxed{\text{Cut}^{(2)}_{\nb{}}\left[\mathcal{M}_4^{\text{1-loop}}\right]- \text{Cut}^{(2)}_{\cb{}}\left[\mathcal{M}_4^{\text{1-loop}}\right] +\frac{\sigma}{2m_1 m_2(\sigma^2-1)}\text{Cut}^{(0)}_{\cb{}}\left[\mathcal{M}_4^{\text{1-loop}}\right] = 0.}
\end{equation}
Note that this is a necessary but not a sufficient condition for a quantum LRL symmetry, we must separately impose the cancellation of the power-law contributions.
\begin{table}[t]
    \centering
    \begin{tabular}{| Sc | Sc |}
        \hline
        $b_0$ & $\displaystyle\frac{1}{8}\left[ \kappa^2m_1^2m_2^2\left( 16Q + \mathcal{N}\,\text{Re}(gY_L + gY_R) \right) - 32QS \right]$ \\
        \hline
        $b_1$ & $\displaystyle\frac{m_1 m_2}{16}\left[ \kappa^4 m_1^2m_2^2(3\mathcal{N}-14)+24\kappa^2S+8\left(32Q^2+\frac{S^2}{m_1^2 m_2^2}-8(|gY_L|^2+|gY_R|^2)\right) \right]$ \\
        \hline
        $b_2$ & $\displaystyle-\frac{3\kappa^2m_1^2 m_2^2}{8}\left[ \mathcal{N}\,\text{Re}(gY_L + gY_R)+32Q \right]$ \\
        \hline
        $b_3$ & $\displaystyle\frac{m_1 m_2}{16}\left[ \kappa^2 m_1^2m_2^2(40-7\mathcal{N}) - 16\kappa^4 S + 64(|gY_L|^2+|gY_R|^2-2Q^2) \right]$ \\
        \hline
        $b_4$ & $\displaystyle\frac{\kappa^2 m_1^2 m_2^2}{4}\left[ \mathcal{N}\,\text{Re}(gY_L + gY_R)+32Q \right]$ \\
        \hline
        $b_5$ & $\displaystyle \frac{\kappa^4 m_1^3 m_2^3}{4}(\mathcal{N}-6)$ \\
        \hline
    \end{tabular}
    \caption{Explicit expression for the coefficients $b_i$ in \eqref{eq:LHS} for the first quantum correction to the potential. Here, we have defined $gY_{L} = gY_{1,L}gY_{2,L}^*$, $gY_{R} = gY_{1,R}gY_{2,R}^*$, $Q = Q_1 Q_2$ and $S=g_1g_2$.}
    \label{tab:coeff quBox}
\end{table}
In a generic EFT, this combination of cuts is proportional to a fifth degree polynomial in $\sigma$
\begin{equation}
\label{eq:LHS}
    \text{LHS of \eqref{eq:qu arccosh condition}} = -\frac{1}{\sigma^2-1}\sum_{k=0}^5 b_k \sigma^k,
\end{equation}
where the coefficients $b_k$ only depend on the three-point couplings of the massless mediators to the massive states, and their explicit form is recorded in Table \ref{tab:coeff quBox}. Importantly, the leading order in the ultra-relativistic limit $\sigma\to\infty$ only receives contributions from the double graviton and gravitino exchange, leading to the following \textit{universal} prediction for any model of perturbative quantum gravity
\begin{equation}
\label{eq:N=6 constraint}
    \mathcal{M}_4^{\text{1-loop}}\biggr\vert_{\mathcal{O}(\hbar)} \sim 16\left(\mathcal{N}-6\right)G^2 m_1^2 m_2^2 \sigma^2 \log(\sigma)\log(-q^2), \hspace{5mm} \text{as} \hspace{5mm} \sigma\rightarrow \infty,
\end{equation}
where $\mathcal{N}$ denotes the total number of spin-3/2 gravitinos present in the model. As is well-known, the presence of $\mathcal{N}$ species of massless spin-3/2 particles requires that the scattering amplitudes of the entire model satisfy the $\mathcal{N}$-extended supersymmetry Ward identities \cite{Grisaru:1977kk}. This symmetry (or equivalently gauge invariance for the spin-3/2 fields) fixes the cubic couplings involving the gravitinos completely in terms of the gravitational coupling $\kappa$; therefore only the number of species appears as a free parameter in the expression for $b_5$. We refer the reader to Appendix \ref{app:gravitino details} for a more detailed account of the gravitino exchange computation.

Remarkably, the only way for this contribution to vanish is if it includes \textit{exactly} 6 gravitinos. In other words, \textit{$\mathcal{N}=6$ SUGRA is the only possible theory that could cancel the first quantum correction to the bound-state potential}, and thus recover the hydrogen-like degeneracy of the energy levels. The fact that the correction is not canceled instead in a maximally supersymmetric theory is certainly surprising. Nevertheless, this observation appears to be consistent with the analysis of Regge scattering in massless $\mathcal{N}=6$ SUGRA \cite{Bartels:2012ra,Bern:2020gjj}. In Section \ref{subsec:N8sugra}, we explicitly check that the potential in $\mathcal{N}=8$ SUGRA does acquire a correction at $\oh(\hbar)$ in the $\sigma\to\infty$ limit, in agreement with the prediction \eqref{eq:N=6 constraint}.

Although the condition on the number of gravitinos is a necessary requisite to cancel the first quantum correction in the high energy limit, it is not clear whether there is an actual theory of $\mathcal{N}=6$ supergravity in which the 1-loop quantum correction to the potential completely vanishes. In fact, if one imposes that all coefficients $b_k$ are individually zero, an inconsistent condition on the couplings is reached. This is true even when including multiple flavors of massless mediators and independent left- and right-handed couplings for fermionic interactions. In other words, when we restrict ourselves to theories with the interactions enumerated in Tables \ref{tab:3pt massive class} and \ref{tab:3pt massless class}, there is no value of the couplings for which the $\text{arccosh}(\sigma)$ contribution to the quantum potential cancels.

Let us understand why this happens. Using the definition \eqref{eq:LHS}, one can see that massless mediators of different spins $s$ contribute to the numerator polynomial at most at $\mathcal{O}(\sigma^{3+\lceil s\rceil})$. This allows the fifth, fourth and third powers of $\sigma$ to be canceled by tuning the new couplings that enter at these orders. Since the $\sigma^2$ coefficient $b_2$ is proportional to $b_4$, $b_2=0$ does not place any further conditions. Finally, we have $b_1$ and $b_0$ which can be solved simultaneously, but yield a solution with $Q^2<0$. Since on physical grounds, we expect that the charge $Q$ must be real, we conclude that the conditions $b_i=0$ are inconsistent. A corollary of this analysis is that there is a physical choice of couplings for which the amplitude has the best possible high energy behavior of $\mathcal{O}(\sigma^{-1}\log(\sigma))$ when $\sigma\to\infty$.

Nevertheless, we have explicitly omitted certain interactions in our effective Lagrangian such as the inclusion of non-abelian gauge fields or magnetic charges for the massive scalar lines. Thus, we cannot definitively claim that all possible quantum field theories will result in a non-vanishing 1-loop correction to the quantum potential. Our statement is that a theory that recovers the quantum LRL symmetry at 1-loop is necessarily a $\mathcal{N}=6$ supergravity theory, and needs to include a broader class of interactions than the set we consider here. In particular, a simple Kaluza-Klein reduction like the one in Section \ref{sec:Nge4} does not yield an $\mathcal{N}=6$ amplitude that satisfies \eqref{eq:qu arccosh condition}. We leave a more complete characterization of the amplitude in $\mathcal{N}=6$ SUGRA for future work.

\subsection{Quantum degeneracy breaking in \texorpdfstring{$\mathcal{N}=8$}{N=8} SUGRA}
\label{subsec:N8sugra}
In \cite{Caron-Huot:2018ape}, it was shown that $\mathcal{N}=8$ supergravity conserves a relativistic version of the LRL vector at 1-loop. Specifically, the classical orbits of BPS-$\overline{\text{BPS}}$ two-body bound states were shown to have no-precession at 2PM.\footnote{In this section we restrict to the simple ``one-angle'' case \cite{Caron-Huot:2018ape} of electrically charged BPS states with maximal misalignment of charge vectors $\phi_1=\phi_2=-\phi_3=-\phi_4=\pi$.} The authors of \cite{Caron-Huot:2018ape} conjectured that this extends to an exact quantum-mechanical statement of the degeneracy of energy levels. In this section we show explicitly that this conjecture is false.

The complete 1-loop amplitude in $\mathcal{N}=8$ supergravity can be calculated from the four-particle cuts of the box and cross-box integrals, as there are no-triangle or bubble contributions in this theory. Moreover, since these cuts are directly related to the ones in $\mathcal{N}=4$ super-Yang-Mills theory via double copy, we can compute them in the gauge theory and square them to obtain their gravity counterpart. For details on this cut calculation, we refer the reader to Appendix \ref{app:quad}. The final result for the 1-loop amplitude in $\mathcal{N}=8$ SUGRA is
\begin{equation}
    \label{eq:N8}
    \mathcal{M}_4^{\mathcal{N}=8\;\text{SUGRA}}=-\kappa^4 m_1^4 m_2^4\left(\sigma+1+\frac{q^2}{2m_1 m_2}\right)^4\left(  I_{\nb{}} +I_{\cb{}}\right)\,.
\end{equation}
In particular, the relevant soft expanded cuts of the box and crossed box integrals are given by 
\begin{align}
\text{Cut}^{(0)}_{\nb{}}\left[\mathcal{M}_4^{\mathcal{N}=8}\right] &=  -\kappa^4 m_1^4 m_2^4 (\sigma+1)^4\,, \hspace{8mm} \text{Cut}^{(2)}_{\nb{}}\left[\mathcal{M}_4^{\mathcal{N}=8}\right] = -2\kappa^4 m_1^3 m_2^3(\sigma + 1)^3\,,  \nonumber\\
\text{Cut}^{(0)}_{\cb{}}\left[\mathcal{M}_4^{\mathcal{N}=8}\right] &= -\kappa^4 m_1^4 m_2^4 (\sigma+1)^4\,, \hspace{8mm} \text{Cut}^{(2)}_{\cb{}}\left[\mathcal{M}_4^{\mathcal{N}=8}\right] = -2\kappa^4 m_1^3 m_2^3(\sigma + 1)^3\,.
\end{align}
From this, it is clear that the condition \eqref{eq:qu arccosh condition} is violated
\begin{align}
    \label{eq:N8cut} &\text{Cut}^{(2)}_{\nb{}}\left[\mathcal{M}_4^{\mathcal{N}=8}\right]- \text{Cut}^{(2)}_{\cb{}}\left[\mathcal{M}_4^{\mathcal{N}=8}\right] +\frac{\sigma}{2m_1 m_2(\sigma^2-1)}\text{Cut}^{(0)}_{\cb{}}\left[\mathcal{M}_4^{\mathcal{N}=8}\right] = -\frac{\kappa^4 m_1^3 m_2^3}{2} \frac{\sigma(\sigma+1)^3}{\sigma-1}\,.
\end{align}
This result of course follows from the observation, made in Section \ref{subsec:gravitino}, that the condition \eqref{eq:qu arccosh condition} can only hold in a gravitational theory with $\mathcal{N}=6$ supersymmetry. In particular, in the $\sigma\to\infty$ limit, we can use the definition \eqref{eq:LHS} and \eqref{eq:N8cut} to identify 
\begin{align}
    b^{\mathcal{N}=8}_5 = \frac{\kappa^4 m_1^3 m_2^3}{2}\,.
\end{align}
This exactly matches our result for $b_5$ given in Table \ref{tab:coeff quBox} upon setting $\mathcal{N}=8$, providing a cross-check of the two different methods used in this section (on-shell superspace) and the previous one (summing component amplitudes). 

We now extend this result by calculating the explicit quantum potential and subsequent correction to the energy levels. The result for the quantum correction to the potential is
\begin{align}
    V_{\mathcal{N}=8}(\mathbf{x},\mathbf{p})\biggr\vert_{\mathcal{O}(\hbar)} &=  -\frac{4G^2 m_1^2 m_2^2}{\pi  r^3 }\frac{\left(\sigma+1\right)^4 \left(\sqrt{\sigma ^2-1} -\sigma\text{arccosh}\left(\sigma\right)\right)}{E_1(\mathbf{p}) E_2(\mathbf{p})\left(\sigma ^2-1\right)^{3/2}} \nonumber\\
    &\overset{\sigma \rightarrow 1}{=} \frac{64}{3\pi}\frac{G^2 m_1 m_2}{ r^3}\,.
\end{align}
From the static limit of this potential we calculate the leading breaking of the LRL degeneracy 
\begin{equation}
    \delta E_{n,l}^{\mathcal{N}=8} = \frac{64G^2 m_1 m_2}{3\pi} \biggr\langle \frac{1}{r^3}\biggr\rangle = \frac{32768 G^5 m_1^7 m_2^7}{3\pi (m_1+m_2)^3 l\left(l+\frac{1}{2}\right)(l+1)n^3}\,,
\end{equation}
where the effective Bohr radius in this model is $a_0\vert_{\mathcal{N}=8} = \frac{m_1+m_2}{8G m_1^2 m_2^2}$.

\section{Discussion}
\label{sec:discussion}

In this paper, we have derived the general conditions for the presence of a hidden LRL symmetry in gravitational two-body bound states at second post-Minkowskian order. The conditions have simple statements in terms of the cancellation of certain integral coefficients of the corresponding 1-loop scattering amplitudes, both classically (\ref{eq:tricut}) and quantum-mechanically (\ref{eq:qu arccosh condition}). We have presented the classical analysis in the context of a maximally generic EFT of long-range interactions mediated by scalar, photon and graviton mediators with general interactions, and generalized this for the quantum analysis to include massless fermion mediators and internal spin-transitions. The generality of our setup allows us to draw some powerful and surprising conclusions. Despite the classical no-precession condition \eqref{eq:tricut} being satisfied by a broad class of models (e.g.\@ $\mathcal{N}\ge4$ supergravities), preserving LRL symmetry at first quantum order is a much more constraining property. We find that solving the quantum degeneracy condition \eqref{eq:qu arccosh condition} in the ultra-relativistic limit requires the existence of exactly six Majorana gravitinos, therefore only a model of $\mathcal{N}=6$ supergravity can preserve LRL symmetry at first quantum order.

There are several novel aspects to our calculations that we would like to highlight. Firstly, the extension of on-shell PM techniques to quantum corrections, such as matching the gravitational scattering amplitude to a bound-state potential at all-orders-in-velocity \cite{Cheung:2018wkq}. This opens up several avenues for future research. For example, it would be interesting to study quantum corrections to different observables such as Love numbers or gravitational waveforms, perhaps relating them to the recent results obtained in the semiclassical regime \cite{Aoude:2024sve,Aoude:2025jvt,Aoki:2025ihc,Akpinar:2026oni,Ilderton:2025aql,Carrasco:2025bgu}. 

The second novel aspect is the inclusion of the exchange of pairs of massless fermionic mediators and the resulting quantum contributions to the effective potential. We have shown that in the ultra-relativistic limit of scattering, $\sigma\rightarrow \infty$, graviton and gravitino contributions appear at the same order $\sim\sigma^2 \log(\sigma)$ and can cancel exactly only for $\mathcal{N}=6$. It is left to future work to understand the significance of this observation and determine whether similar improved UV behavior continues at higher loops. In addition there may be other contexts where the inclusion of massless fermion exchange contributions to long-range interactions leads to interesting new results. For example there is an extensive phenomenological literature on \textit{neutrino forces} in the Standard Model \cite{Feinberg:1968zz,Hsu:1992tg,Ghosh:2024qai,Ghosh:2024ctv}. Most of these calculations are restricted to the calculation of the static potential, it may be of interest to apply the ``post-Minkowskian'' EFT matching approach developed in this paper to this context and calculate velocity-resummed expressions for the two-neutrino potential. 

Crucially, the approach taken in this paper of calculating corrections to the quantum potential order-by-order in $\hbar$ assumes the existence of a sensible semi-classical expansion. It would be very interesting to extend this kind of generic EFT analysis to a more general class of intrinsically quantum-mechanical models, including non-abelian gauge theories. It should be possible to reproduce, and possibly generalize, the claim of LRL degeneracy in bound states of $\mathcal{N}=4$ SYM \cite{Caron-Huot:2014gia}. A related important question is whether there is a quantum analogue of the classical radial action, from which one could directly obtain different observables such as the relativistic energy levels. The quantum version of such a boundary-to-bound dictionary may need to be phrased as a connection between the scattering amplitude and the Bethe-Salpeter kernel \cite{Salpeter:1951sz,Adamo:2022ooq}. This would allow one to bypass matching to a potential at each order in the velocity expansion and instead perform the calculation fully relativistically.

We would also like to highlight some extensions of our results that have been left to future work. In the case of single-flavor scalar/vector mediators, the classical no-precession condition is satisfied by specific Kaluza-Klein-like solutions such as the black-hole $a$-model \cite{Ortin:2004ms,deNeeling:2023egt}. These are characterized by the presence of a scalar field that is coupled non-minimally via an exponential gauge kinetic function. Our 1-loop calculation reproduces the linear term in an expansion in small coupling, but one should be able to match the complete exponential behavior by imposing the no-precession condition at all orders in $G$ in the probe limit. It could be interesting to extend these higher-loop calculations beyond the probe limit in a self-force expansion using the frameworks developed in \cite{Kosmopoulos:2023bwc, Cheung:2023lnj}.

 As is evident from the explicit 1-loop amplitudes calculated in this paper, the high-energy limit $\sigma\to\infty$ of the first quantum correction is badly behaved, scaling generically as $\sim \sigma^2\log(\sigma)$. This is similar to the behavior of the classical amplitude at 3PM order in the classical regime, where it cancels with radiation contributions in physical observables, leading to a smooth transition to the massless limit \cite{Herrmann:2021tct}. Due to the absence of radiation at 2PM we expect that the quantum log has a qualitatively different resolution. It would be interesting to understand in detail the nature of the transition between high-energy massive and massless Regge scattering \cite{Bartels:2012ra,Bern:2020gjj,Rothstein:2024nlq,Alessio:2025isu,Saavedra:2026ttk}. 

Finally, in this paper we have not identified a specific model of massive particles coupled to $\mathcal{N}=6$ SUGRA for which the quantum-degeneracy constraint is satisfied for all values of $\sigma$. As explained in Section \ref{subsec:gravitino}, there is an obstruction at sub-leading orders in the ultra-relativistic expansion, and given the EFT parametrization used in this paper there are \textit{no physical choices of couplings that complete the cancellation}. Nonetheless we are hesitant to claim a strict \textit{no-go result} for LRL degeneracy in quantum gravity; there are possibly other interactions that we have not included that may be important. For example, we exclude massive states with magnetic or more exotic charges which, as suggested by results in certain Taub-NUT theories \cite{Guevara:2024edh}, could satisfy the necessary conditions. Indeed given that the leading constraint in the high-energy limit already restricts us to $\mathcal{N}=6$ SUGRA, the question of satisfying the quantum-degeneracy condition is probably best addressed within the framework of the double copy and on-shell superamplitudes \cite{Herderschee:2019ofc,Herderschee:2019dmc}. We leave this analysis for future work.

\vspace{3mm}
\noindent \textbf{Acknowledgments}

\vspace{3mm}
We would like to thank Emil Bjerrum-Bohr, John Joseph Carrasco, Johannes Henn, Aidan Herderschee, Scott Melville, Michael Ruf and Michael Saavedra for helpful discussions. CRTJ is supported by funding from the University of Arizona. SP and MS are supported by the US Department of Energy under contract DE-SC0010010 Task F and by Simons Investigator Award
\#376208 (SP). CRTJ would like to thank Brown University for hospitality during various stages of preparation of this work. 

\appendix

\newpage

\section{Quantum Soft Contributions to Eikonal Integrals}
\label{app:integrals}

To calculate the quantum corrections to the bound-state energies in Section \ref{sec:quantum} we require the evaluation of 1-loop \textit{eikonal} or \textit{soft} integrals. The evaluation of the potential region contribution to these integrals, relevant for classical PM calculations, is now well-known \cite{Parra-Martinez:2020dzs}. To calculate the quantum corrections we need the result in the full soft region (including contributions from the so-called \textit{quantum soft} region, defined below).

The 1-loop integral contributions to long-range interactions reviewed in Section \ref{subsec:reviewclassical} are reducible to the well-known basis of scalar integrals
\begin{equation}
\label{eq:fullint}
    \begin{aligned}
    I_{\bub{}} &:= \int  \frac{\hat{\text{d}}^{D}\ell}{[\ell^2+i0] [(\ell-q)^2+i0]}\\
    I_{\tric{}} &:= \int  \frac{\hat{\text{d}}^{D}\ell}{[\ell^2+i0] [(\ell-q)^2+i0] [(p_1+\ell)^2-m_1^2+i0]}\\
    I_{\tricinv{}} &:= \int  \frac{\hat{\text{d}}^{D}\ell}{[\ell^2+i0] [(\ell-q)^2+i0] [(p_2-\ell)^2-m_2^2+i0]}\\
    I_{\nb{}} &:= \int  \frac{\hat{\text{d}}^{D}\ell}{[\ell^2+i0] [(\ell-q)^2+i0] [(p_1+\ell)^2-m_1^2+i0][(p_2-\ell)^2-m_2^2+i0]}\\
    I_{\cb{}} &:= \int  \frac{\hat{\text{d}}^{D}\ell}{[\ell^2+i0] [(\ell-q)^2+i0] [(p_1+\ell)^2-m_1^2+i0][(p_2+q+\ell)^2-m_2^2+i0]}.
\end{aligned}
\end{equation}
To expand these integrals in the soft limit (\ref{eq:softexpansion}) it is convenient to define a new set of \textit{barred} variables
\begin{equation}
    \overline{m}_1 \overline{u}_1 := p_1+\frac{1}{2}q, \hspace{10mm}  \overline{m}_2 \overline{u}_2 := p_2 - \frac{1}{2}q, \hspace{10mm} \bar m_i := \sqrt{m_i^2 - \frac{1}{4}q^2},
\end{equation}
where the external kinematic variables satisfy
\begin{equation}
    \overline{u}_1 \cdot \overline{u}_2 := y, \hspace{10mm} \overline{u}_i \cdot q =  0, \hspace{10mm} \overline{u}_i^2 = 1.
\end{equation}
The full integrals (\ref{eq:fullint}) are then expanded in a barred version of the soft region
\begin{equation}
    \ell^\mu \sim q^\mu \ll \overline{u}_i^\mu,
\end{equation}
the result is expressible in terms of the following integral family\footnote{The superscript $\pm$ in (\ref{eq:Gdef}) is irrelevant for all of the integrals except for $G^\pm_{1,1,1,1}$ where it distinguishes the linearized box and crossed-box integrals; these integrals satisfy the same differential equations but have different boundary conditions. For the bubble and triangle we will suppress this superscript and note that the triangle integrals are trivially related by $G_{1,1,1,0}:= G^{\pm}_{1,1,1,0} = G^{+}_{1,1,0,1} = -G^{-}_{1,1,0,1}$.}
\begin{equation}
    \label{eq:Gdef}
    G^\pm_{a_1,a_2,a_3,a_4} := \int  \frac{\hat{\text{d}}^{D}\ell}{[\ell^2+i0]^{a_1} [(\ell-q)^2+i0]^{a_2} [\overline{u}_1\cdot \ell+i0]^{a_3}[\overline{u}_2\cdot \ell\pm i0]^{a_4}}.
\end{equation}
After calculating the required $G$-integrals, we then re-expand in the original soft expansion to the desired order using
\begin{equation}
    y = \frac{m_1 m_2 \sigma + \frac{1}{4}q^2}{\sqrt{\left( m_1^2 - \frac{1}{4}q^2 \right)\left( m_2^2 - \frac{1}{4}q^2 \right)}} = \sigma + \frac{\sigma(m_1^2+m_2^2)+2m_1 m_2}{8m_1^2 m_2^2}q^2+ \oh(q^4),
\end{equation}
where $\sigma:=\frac{p_1\cdot p_2}{m_1 m_2}$. The integration-by-parts system for the family (\ref{eq:Gdef}) has 4 master integrals: the \textit{bubble} $G_{1,1,0,0}$, the \textit{triangles} $G_{1,1,1,0}$ and $G_{1,1,0,1}$ and the \textit{box} $G_{1,1,1,1}$. We calculate these integrals by the method of differential equations \cite{Henn:2013pwa}. Applying the differential operator 
\begin{equation}
    \frac{\partial}{\partial y} = \frac{1}{y^2-1}\left(y \overline{u}_1^\mu - \overline{u}_2^\mu\right) \frac{\partial}{\partial \overline{u}_1^\mu},
\end{equation}
to the master integrals and integration-by-parts reducing the result, we find the following closed differential equation system 
\begin{equation}
    \label{eq:DE}
    \frac{\partial}{\partial y}\begin{pmatrix}
        G_{1,1,0,0} \\
        G_{1,1,1,0} \\
        G_{1,1,0,1} \\
        G_{1,1,1,1}
    \end{pmatrix} = \begin{pmatrix}
        0 & 0 & 0 & 0\\
        0 & 0 & 0 & 0\\
        0 & 0 & 0 & 0\\
        \frac{4(D-3)}{q^2(y^2-1)} & 0 & 0 & -\frac{y}{y^2-1}
    \end{pmatrix}\begin{pmatrix}
        G_{1,1,0,0} \\
        G_{1,1,1,0} \\
        G_{1,1,0,1} \\
        G_{1,1,1,1}
    \end{pmatrix}.
\end{equation}
The bubble integral is a standard result 
\begin{equation}
    \label{eq:bubble}
    G_{1,1,0,0} = \frac{i}{(4\pi)^{\frac{D}{2}}} \frac{\Gamma\left(\frac{4-D}{2}\right) \Gamma\left(\frac{D-2}{2}\right)^2}{\Gamma\left(D-2\right)} \left(-q^2\right)^{\frac{D-4}{2}}.
\end{equation}
The triangle integral can be obtained by direct integration
\begin{align}
    \int  \frac{\hat{\text{d}}^{D}\ell}{[\ell^2+i0] [(\ell-q)^{2}+i0] [\overline{u}_1\cdot \ell+i0]} &= \frac{1}{2} \int  \frac{\hat{\text{d}}^{D}\ell}{[\ell^2+i0] [(\ell-q)^{2}+i0]} \left[\frac{1}{\overline{u}_1\cdot \ell+i0} - \frac{1}{\overline{u}_1\cdot \ell-i0}\right]\nonumber\\
    &=  -\frac{i}{2}\int  \hat{\text{d}}^{D}\ell \frac{\hat{\delta}(\overline{u}_1\cdot \ell)}{[\ell^2+i0] [(\ell-q)^{2}+i0]},
\end{align}
where in the second expression we have made a strategic change of variables $\ell\rightarrow q-\ell$ and used the Sokhotski–Plemelj theorem to write the result as a cut integral. We can evaluate this remaining integral in the rest-frame of body-1 
\begin{equation}
    \label{eq:rest1kin}
    \overline{u}_1^\mu = \left(1,\mathbf{0}\right), \hspace{5mm} \overline{u}_2^\mu = \left(\sqrt{1+\mathbf{v}^2},\mathbf{v}\right), \hspace{5mm} q^\mu = \left(0,\mathbf{q}\right),
\end{equation}
where $\mathbf{v}\cdot \mathbf{q}=0$, $\mathbf{q}^2 = -q^2$ and $y = \sqrt{1+\mathbf{v}^2}$. The above then reduces to a $(D-1)$-dimensional Euclidean bubble integral 
\begin{align}
    \int  \frac{\hat{\text{d}}^{D}\ell}{[\ell^2+i0] [(\ell-q)^{2}+i0] [\overline{u}_1\cdot \ell+i0]} &= -\frac{i}{2} \int \hat{\text{d}}\omega \hat{\text{d}}^3\boldsymbol{\ell}\frac{\hat{\delta}(\omega)}{[\omega^2-\boldsymbol{\ell}^2+i0] [\omega^2-(\boldsymbol{\ell}-\mathbf{q})^2+i0]}\nonumber\\
    &=  -\frac{i}{2} \int \frac{\hat{\text{d}}^3\boldsymbol{\ell}}{\boldsymbol{\ell}^2 (\boldsymbol{\ell}-\mathbf{q})^2},
\end{align}
and therefore
\begin{align}
    G_{1,1,1,0} = -\frac{i}{2(4\pi)^{\frac{D-1}{2}}} \frac{\Gamma\left(\frac{5-D}{2}\right) \Gamma\left(\frac{D-3}{2}\right)^2}{\Gamma\left(D-3\right)} \left(-q^2\right)^{\frac{D-5}{2}}.
\end{align}
For the box integral we can straightforwardly write down the general solution of the differential equation, treating the bubble integral (\ref{eq:bubble}) as a ``source''
\begin{equation}
    \label{eq:solDE}
    G^{\pm}_{1,1,1,1}(y) = \frac{c_{\pm}(q^2)}{\sqrt{y^2-1}} +\frac{4(D-3)\text{arccosh}\left(y\right)}{q^2 \sqrt{y^2-1}} G_{1,1,0,0},
\end{equation}
where $c_\pm(q^2)$ is an integration constant determined by imposing a suitable boundary condition in the static limit $y\rightarrow 1$. This can be calculated systematically using the method of regions \cite{Beneke:1997zp}. We analyze the integral using the rest-frame kinematics (\ref{eq:rest1kin}); explicitly, the non-relativistically parametrized linearized box integral has the form 
\begin{equation}
    G^{\pm}_{1,1,1,1} = \int  \frac{\hat{\text{d}} \omega \;\hat{\text{d}}^3\boldsymbol{\ell}}{[\omega^2-\boldsymbol{\ell}^2+i0] [\omega^2-(\boldsymbol{\ell}-\mathbf{q})^2+i0][\omega+i0][\sqrt{1+\mathbf{v}^2}\omega-\mathbf{v}\cdot \boldsymbol{\ell} \pm i0]}.
\end{equation}
In the limit $\mathbf{v}\rightarrow 0$ we expect that this integral may have contributions from three different regions:\footnote{Physically, while the potential region corresponds to \textit{off-shell} modes that mediate instantaneous-in-time interactions, both the radiation and quantum soft regions describe the exchange of \textit{on-shell} radiation modes. The quantum soft radiation modes have energy of order the soft scale $\omega \sim |\mathbf{q}|$ and are therefore considered ``heavy'' and integrated out, while the radiation region modes have energies of order the so-called ultra-soft scale $\omega \sim |\mathbf{v}||\mathbf{q}|\ll |\mathbf{q}|$ and therefore remain as propagating degrees-of-freedom in the potential EFT.}
\begin{enumerate}
    \item \textit{potential region}: \hspace{11mm} $(\omega,\boldsymbol{\ell}) \sim |\mathbf{q}|(\mathbf{v},1)$ 
    \item \textit{radiation region}: \hspace{10mm} $(\omega,\boldsymbol{\ell}) \sim |\mathbf{q}|(\mathbf{v},\mathbf{v})$ 
    \item \textit{quantum soft region}: \hspace{4mm} $(\omega,\boldsymbol{\ell}) \sim |\mathbf{q}|(1,1)$. 
\end{enumerate}
By inspection, we can determine the leading-order contribution of each region
\begin{equation}
    G^{\pm}_{1,1,1,1}\biggr\vert_{\text{pot}} \sim \mathbf{v}^{-1}, \hspace{10mm} G^{\pm}_{1,1,1,1}\biggr\vert_{\text{rad}} \sim \mathbf{v}^{D-4}, \hspace{10mm} G^{\pm}_{1,1,1,1}\biggr\vert_{\text{q-soft}} \sim \mathbf{v}^{0}.
\end{equation}
It is trivial to show that, at 1-loop, order-by-order the contributions from the radiation region are scaleless integrals; the full soft integrals are therefore expected to be the sum of the potential and quantum soft region contributions. We conclude that the potential region gives the dominant contribution in the limit $\mathbf{v} \rightarrow 0$ and we can use that as a boundary condition. Explicitly we need to calculate 
\begin{equation}
    G^{\pm}_{1,1,1,1}\biggr\vert_{\mathcal{O}(\mathbf{v}^{-1})} = \int  \frac{\hat{\text{d}} \omega \;\hat{\text{d}}^3\boldsymbol{\ell}}{[\boldsymbol{\ell}^2] [(\boldsymbol{\ell}-\mathbf{q})^2][\omega+i0][\omega-\mathbf{v}\cdot \boldsymbol{\ell} \pm i0]}.
\end{equation}
Unlike higher-order terms in the velocity expansion, the $\omega$-integral in this expression is well-defined without further regularization and can be evaluated using Cauchy's theorem. For $G^{-}_{1,1,1,1}$ we close the contour in the lower half-plane (LHP) and pick up the residue at $\omega = -i0$, the result reduces to a $(D-2)$-dimensional Euclidean bubble integral. The boundary condition in this case is then
\begin{equation}
    \label{eq:boxbc}
    G^-_{1,1,1,1}\biggr\vert_{\mathcal{O}(\mathbf{v}^{-1})} = \frac{ \Gamma
   \left(\frac{6-D}{2}\right) \Gamma
   \left(\frac{D-4}{2}\right)^2}{2(4 \pi )^{\frac{D-2}{2}}\Gamma (D-4)} \frac{\left(-q^2\right)^{\frac{D-6}{2}}}{\sqrt{y^2-1}}.
\end{equation}
For $G^{+}_{1,1,1,1}$ there are no $\omega$-residues in the LHP and so the $\mathcal{O}\left(\mathbf{v}^{-1}\right)$ contribution vanishes 
\begin{equation}
    \label{eq:cboxbc}
    G^{+}_{1,1,1,1}\biggr\vert_{\mathcal{O}(\mathbf{v}^{-1})}  = 0. 
\end{equation}
This result can be anticipated by noting that the full crossed-box integral is not expected to have an $s$-channel discontinuity.

These boundary conditions are sufficient to determine the integration constants $c_\pm(q^2)$ in (\ref{eq:solDE}). Expanding in $D=4-2\epsilon$ and giving only the contributions that are non-analytic in $q^2$ we arrive at the final results 
\begin{align}
    \label{eq:G1100}
    G_{1,1,0,0} &= -\frac{i}{16 \pi ^2} \log\left(-q^2\right),\\
    \label{eq:G1110}
    G_{1,1,1,0} &= -\frac{i}{16 \sqrt{-q^2}},\\
    \label{eq:G1111plus}
    G^{+}_{1,1,1,1} &= -\frac{i}{4\pi^2\sqrt{y^2-1}}\text{arccosh}\left(y\right)\frac{\log(-q^2)}{q^2},\\
    \label{eq:G1111minus}
    G^{-}_{1,1,1,1} &= -\frac{i}{4\pi^2\sqrt{y^2-1}}\left[ \text{arccosh}\left(y\right)-i\pi\right]\frac{\log(-q^2)}{q^2}.
\end{align}
There are a few sanity checks of these results. The full (not linearized) box integral is given as equation (4.40) in \cite{Ellis:2007qk}. Identifying kinematic variables $s_{12} \rightarrow q^2$, $s_{23} \rightarrow m_1^2 +m_2^2 +2m_1 m_2 y$ and expanding to leading order in small $q$ gives exactly (\ref{eq:G1111minus}). Additionally, an appropriate linear combination of the box and crossed-box integrals simplifies to a cut box that can be directly evaluated 
\begin{align}
    G^-_{1,1,1,1} - G^+_{1,1,1,1} &= -\frac{i}{2}\int \hat{\text{d}}^{D}\ell \frac{\hat{\delta}(\overline{u}_1\cdot \ell)\hat{\delta}(\overline{u}_2\cdot \ell)}{[\ell^2+i0] [(\ell-q)^2+i0]} \nonumber\\
    &= - \frac{1}{4\pi \sqrt{y^2-1}} \frac{\log(-q^2)}{q^2},
\end{align}
where again we are displaying only non-analytic terms. This result is clearly satisfied by (\ref{eq:G1111plus}) and (\ref{eq:G1111minus}). Alternatively, assuming the regularity of the crossed-box integral $G^+_{1,1,1,1}$ in the static limit (\ref{eq:cboxbc}), which is independent of the region expansion, this relation can be used as a cross-check of the boundary condition for the box integral $G^-_{1,1,1,1}$ (\ref{eq:boxbc}).

The expressions (\ref{eq:G1100}-\ref{eq:G1111minus}) are the integrals in the full relativistic soft region, that is the sum of the potential and quantum soft non-relativistic regions; it may be of some interest to consider these contributions independently. The bubble integral $G_{1,1,0,0}$ vanishes in the potential region, and therefore the general solution of the differential equation (\ref{eq:DE}) is simpler and does not contain an $\text{arccosh}\left(\sigma\right)$. Conversely the triangle integral $G_{1,1,1,0}$ has only a potential region contribution. We therefore find the expected potential region contribution \cite{Parra-Martinez:2020dzs}
\begin{align}
    G_{1,1,0,0}\biggr\vert_{\text{pot}} &= 0,\\
    G_{1,1,1,0}\biggr\vert_{\text{pot}} &= -\frac{i}{16 \sqrt{-q^2}},\\
    G^{+}_{1,1,1,1}\biggr\vert_{\text{pot}} &= 0,\\
    G^{-}_{1,1,1,1}\biggr\vert_{\text{pot}} &= - \frac{1}{4\pi \sqrt{y^2-1}} \frac{\log(-q^2)}{q^2}.
\end{align}
To find the quantum soft contribution we simply take the difference with the full soft region giving 
\begin{align}
    G_{1,1,0,0}\biggr\vert_{\text{q-soft}} &= -\frac{i}{16 \pi ^2} \log\left(-q^2\right),\\
    G_{1,1,1,0}\biggr\vert_{\text{q-soft}} &= 0,\\
    G^{+}_{1,1,1,1}\biggr\vert_{\text{q-soft}} &= -\frac{i}{4\pi^2\sqrt{y^2-1}}\text{arccosh}\left(y\right)\frac{\log(-q^2)}{q^2},\\
    G^{-}_{1,1,1,1}\biggr\vert_{\text{q-soft}} &= -\frac{i}{4\pi^2\sqrt{y^2-1}}\text{arccosh}\left(y\right)\frac{\log(-q^2)}{q^2}.
\end{align}
As expected the box and crossed-box contributions from the quantum soft region are identical, indicating that they are insensitive to the matter propagator $\pm i0$ prescription.

\section{BPS Four-Particle Cuts in \texorpdfstring{$\mathcal{N}=8$}{N=8} SUGRA}
\label{app:quad}

As discussed in Section \ref{subsec:N8sugra}, the scattering amplitudes relevant for the calculation of $\text{BPS}-\overline{\text{BPS}}$ bound-state energy levels in $\mathcal{N}=8$ SUGRA only have box integral contributions at 1-loop. The amplitudes can therefore be fixed completely by calculating a quadruple unitarity cut for each diagram; in this appendix we provide the details of this cut calculation. To do this, we use the massive on-shell superspace formalism, following the notation and conventions introduced in \cite{Herderschee:2019dmc}. The 3-point $\mathcal{N}=8$ SUGRA amplitudes are squares of the ones in $\mathcal{N}=4$ SYM due to the double copy. Thus it suffices to compute the different cuts in the gauge theory and use them to obtain their gravity counterparts.

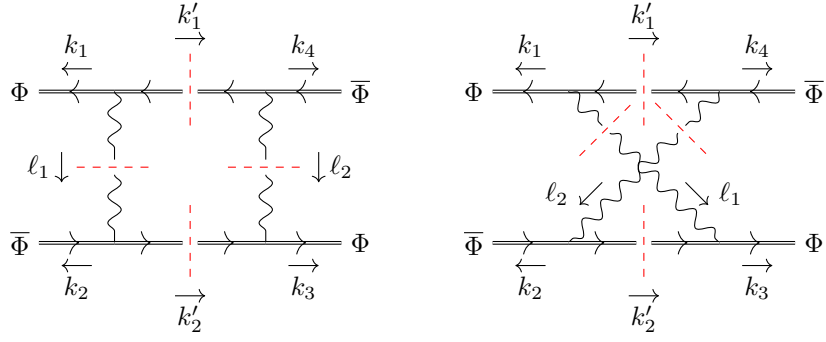
\begin{figure}
    \centering
    \begin{tikzpicture}
    \begin{scope}
    \draw[fermionbar, double] (0,0)--(1,0);
    \draw[fermionbar, double] (1,0)--(1.9,0);
    \draw[fermionbar, double] (2.1,0)--(3,0);
    \draw[fermionbar, double] (3,0)--(4,0);
    \draw[vector] (1,0)--(1,-0.9);
    \draw[vector] (3,0)--(3,-0.9);
    \draw[vector] (1,-1.1)--(1,-2);
    \draw[vector] (3,-1.1)--(3,-2);
    \draw[fermion, double] (0,-2)--(1,-2);
    \draw[fermion, double] (1,-2)--(1.9,-2);
    \draw[fermion, double] (2.1,-2)--(3,-2);
    \draw[fermion, double] (3,-2)--(4,-2);
    \draw[-, color=red, dashed] (2,0.5)--(2,-0.5);
    \draw[-, color=red, dashed] (2,-1.5)--(2,-2.5);
    \draw[-, color=red, dashed] (0.5,-1)--(1.5,-1);
    \draw[-, color=red, dashed] (2.5,-1)--(3.5,-1);   
    \node[left] at (0,0) {$\Phi$};
    \node[left] at (0,-2) {$\overline{\Phi}$};
    \node[right] at (4,0) {$\overline{\Phi}$};
    \node[right] at (4,-2) {$\Phi$};
    \draw[->] (0.7,0.3)--(0.3,0.3);
    \draw[->] (1.8,0.7)--(2.2,0.7);
    \draw[->] (0.7,-2.3)--(0.3,-2.3);
    \draw[->] (1.8,-2.7)--(2.2,-2.7);
    \draw[->] (3.3,0.3)--(3.7,0.3);
    \draw[->] (3.3,-2.3)--(3.7,-2.3);
    \draw[->] (0.3,-0.8)--(0.3,-1.2);
    \draw[->] (3.7,-0.8)--(3.7,-1.2);
    \node at (0.5,0.6) {$k_1$};
    \node at (2,1) {$k'_1$};
    \node at (3.5,0.6) {$k_4$};
    \node at (0.5,-2.6) {$k_2$};
    \node at (2,-3) {$k'_2$};
    \node at (3.5,-2.6) {$k_3$};
    \node at (0,-1) {$\ell_1$};
    \node at (4,-1) {$\ell_2$};
    \end{scope}
    \begin{scope}[xshift=6cm]
    \draw[fermionbar, double] (0,0)--(1,0);
    \draw[fermionbar, double] (1,0)--(1.9,0);
    \draw[fermionbar, double] (2.1,0)--(3,0);
    \draw[fermionbar, double] (3,0)--(4,0);
    \draw[vector] (1,0)--(1.45,-0.45);
    \draw[vector] (1.55,-0.55)--(3,-2);

    \draw[vector] (3,0)--(2.55,-0.45);
    \draw[vector] (2.45,-0.55)--(1,-2);
    \draw[fermion, double] (0,-2)--(1,-2);
    \draw[fermion, double] (1,-2)--(1.9,-2);
    \draw[fermion, double] (2.1,-2)--(3,-2);
    \draw[fermion, double] (3,-2)--(4,-2);
    \draw[-, red, dashed] (2,0.5)--(2,-0.5);
    \draw[-, red, dashed] (2,-1.5)--(2,-2.5);
    \draw[-, red, dashed] (1.15,-0.85)--(1.85,-0.15);
    \draw[-, red, dashed] (2.15,-0.15)--(2.85,-0.85);
    \node[left]  at (0,0)  {$\Phi$};
    \node[left]  at (0,-2) {$\overline{\Phi}$};
    \node[right] at (4,0)  {$\overline{\Phi}$};
    \node[right] at (4,-2) {$\Phi$};
    \draw[->] (0.7,0.3)--(0.3,0.3);
    \draw[->] (1.8,0.7)--(2.2,0.7);
    \draw[->] (3.3,0.3)--(3.7,0.3);

    \draw[->] (0.7,-2.3)--(0.3,-2.3);
    \draw[->] (1.8,-2.7)--(2.2,-2.7);
    \draw[->] (3.3,-2.3)--(3.7,-2.3);
    \draw[->] (2.55,-1.2)--(2.85,-1.5);
    \draw[->] (1.45,-1.2)--(1.15,-1.5);
    \node at (0.5,0.6)  {$k_1$};
    \node at (2,1)      {$k'_1$};
    \node at (3.5,0.6)  {$k_4$};

    \node at (0.5,-2.6) {$k_2$};
    \node at (2,-3)     {$k'_2$};
    \node at (3.5,-2.6) {$k_3$};

    \node at (3.15,-1.35) {$\ell_1$};
    \node at (0.85,-1.35) {$\ell_2$};

\end{scope}
\end{tikzpicture}
    \caption{Box and crossed-box four-particle cuts for $\mathcal{N}=8$ SUGRA.}
    \label{fig:quadcut}
\end{figure}

In $\mathcal{N}=4$ SYM on the Coulomb branch, the $\frac{1}{2}$-BPS multiplets are conveniently encoded in on-shell superfields of the form 
\begin{align}
    \mathcal{W} &= \Phi + \eta^a_I \Psi_a^I -\frac{1}{2}\eta_I^a \eta_J^b\left(\epsilon^{IJ}\Phi_{(ab)} + \epsilon_{ab} W^{(IJ)}\right) + \frac{1}{3}\epsilon_{bc} \eta^b_I \eta^c_J \eta^{Ja} \tilde{\Psi}^I_a + \eta_1^1 \eta_1^2 \eta_2^1 \eta_2^2 \tilde{\Phi}\,, \nonumber\\
    \overline{\mathcal{W}} &= \overline{\tilde{\Phi}} + \eta_a^{\dagger I} \overline{\tilde{\Psi}}^a_I -\frac{1}{2}\eta^{\dagger I}_a \eta^{\dagger J}_b\left(\epsilon_{IJ}\overline{\Phi}^{(ab)} + \epsilon^{ab} \overline{W}_{(IJ)}\right) + \frac{1}{3}\epsilon^{bc} \eta_b^{\dagger I} \eta_c^{\dagger J} \eta^\dagger_{Ja} \overline{\Psi}_I^a + \eta^{\dagger 1}_1 \eta^{\dagger 1}_2 \eta^{\dagger 2}_1 \eta^{\dagger 2}_2\overline{\Phi}\,,
\end{align}
where $I,J$ are $SU(2)_{\text{LG}}$ massive little-group indices and $a,b,c$ are $SU(2)_R$ R-indices. We will also use $\mathcal{G}$ to denote the superfield for the massless vector multiplet states, including the gluon; the details of the components are not important since they are internal but can be found in \cite{Herderschee:2019dmc}.

To calculate the four-particle cuts shown in Figure \ref{fig:quadcut} we require the 3-particle superamplitude $\mathcal{A}_3\left(\mathcal{W}_1,\overline{\mathcal{W}}_2,\mathcal{G}_3\right)$, the explicit form of which is given in equation (5.23) of \cite{Herderschee:2019dmc}. In particular, since we always project one of the massive particles to an (anti-)BPS-scalar $\Phi$ ($\overline{\Phi}$) at the bottom (top) of the supermultiplet, we have
\begin{align}
\mathcal{A}_3\left(\Phi_1,\overline{\mathcal{W}}_2,\mathcal{G}_3\right) &= -g_\text{YM}\frac{x}{m} \prod_{a=1}^2 \left( \frac{1}{2}m (\eta_2^a)^2\eta_3^a + \frac{m}{2x}(\eta_2^a)^2\eta_3^{a+2}-[2^I 3]\eta_{2I}^a\eta_3^a\eta_3^{a+2} \right)\,,\\
\mathcal{A}_3\left(\mathcal{W}_1,\overline{\Phi}_2,\mathcal{G}_3\right) &= -g_\text{YM}\frac{x}{m} \prod_{a=1}^2 \left( [3 2^I]\eta_{2I}^a + \frac{1}{2}m\eta_3^a + \frac{m}{2x}\eta_3^{a+2} \right)\,,
\end{align}
where we have defined
\begin{equation}
    x=\frac{1}{m}\frac{[\xi|2|3\rangle}{[\xi 3]}\quad  \text{ and }\quad (\eta^a_i)^2 = \eta^a_{iI}\eta^{aI}_{i}\,.
\end{equation}
The values for the mass and momenta must be assigned appropriately for each sub-amplitude.

Using this we calculate the box cut with external states projected onto the component R-singlet scalars. For instance, one can first perform the Grassmann integration over the massless mediators, obtaining a product of two 4-point tree-level amplitudes
\begin{align}
\label{eq:cut4int}
&\text{Cut}_{\nb{}}\Big[\mathcal{A}_4\big(\Phi_{k_1},\overline{\Phi}_{k_2},\Phi_{k_3},\overline{\Phi}_{k_4}\big)\Big]\nonumber\\
&=\int \text{d}\eta_{\ell_1} \text{d}\eta_{\ell_2} \text{d}^2\eta_{k'_1}\text{d}^2\eta_{k'_2}\biggr[\mathcal{A}_3\big(\Phi_{k_1},\overline{\mathcal{W}}_{k'_1},\mathcal{G}_{\ell_1}\big)\mathcal{A}_3\left(\mathcal{W}_{k'_2},\overline{\Phi}_{k_2},\mathcal{G}_{-\ell_1}\right)\nonumber\\
&\hspace{40mm}\times \mathcal{A}_3\left(\Phi_{k_3},\overline{\mathcal{W}}_{-k'_2},\mathcal{G}_{-\ell_2}\right)\mathcal{A}_3\left(\mathcal{W}_{-k'_1},\overline{\Phi}_{k_4},\mathcal{G}_{\ell_2}\right)\biggr] \nonumber\\
    &= g_\text{YM}^4\frac{x_1 x_2 x_3 x_4}{m_1^2 m_2^2}\int \text{d}^2\eta_{k'_1}\text{d}^2\eta_{k'_2} \prod_{a=1}^2 \Biggr\{\Bigg[ m_1 m_2\Bigg( \frac{1}{x_2} - \frac{1}{x_1} \Bigg)(\eta_{k_1'}^a)^2 - 2[k_1'^I \ell_1][\ell_1 k_2'^J]\eta_{k_1'\,I}^a\eta_{k_2'\,J}^a \Bigg] \nonumber\\
    &\hspace{55mm}\times\Bigg[ m_1 m_2\Bigg( \frac{1}{x_4} - \frac{1}{x_3} \Bigg)(\eta_{k_2'}^a)^2 - 2[\ell_2 k_1'^I][k_2'^J \ell_2]\eta_{k_1'\,I}^a\eta_{k_2'\,J}^a \Bigg]\Biggr\}\,.
\end{align}
Performing the remaining Grassmann integration over $\eta_{k_1'}$ and $\eta_{k_2'}$ is then equivalent to summing over all possible intermediate massive states. In particular, the fact that \eqref{eq:cut4int} contains terms linear in $\eta_{k_i'}$, indicates that some of these states are fermionic. 

Choosing the reference spinors to be $|\xi_1]=|\xi_2]=|\ell_1]$ and $|\xi_3]=|\xi_4]=|\ell_2]$, writing the spinor products as Dirac traces, and doing some spinor bracket algebra, allows one to obtain the following expression for the box cut
\begin{equation}
    \text{Cut}_{\nb{}}\left[\mathcal{A}_4\left(\Phi_{k_1},\overline{\Phi}_{k_2},\Phi_{k_3},\overline{\Phi}_{k_4}\right)\right] = g_\text{YM}^4\left[(m_1 + m_2)^2 - 2(k_1+ k_3)^2 \right]^2\,.
\end{equation}
This is then mapped to physical scattering kinematics by the substitutions: $k_1 = -p_1$, $k_2 = -p_2$, $k_3= p_2-q$ and $k_4 = p_1+q$, $\ell_1 = -\ell$, $\ell_2 = \ell-q$, which yields
\begin{equation}
\begin{aligned}
\text{Cut}_{\nb{}}\left[\mathcal{A}_4\left(\overline{\Phi}_{-p_1},\Phi_{-p_2},\overline{\Phi}_{p_2-q},\Phi_{p_1+q}\right)\right] &= 4g_\text{YM}^4\left(m_1m_2 + p_1\cdot p_2 +\frac{1}{2}q^2\right) \\&= 4g_\text{YM}^4m_1^2 m_2^2\left(\sigma+1+\frac{q^2}{2m_1 m_2}\right).
\end{aligned}
\end{equation}
Performing a similar computation, the crossed-box cut can be seen to take the same final form
\begin{align}
&\text{Cut}_{\cb{}}\left[\mathcal{A}_4\left(\Phi_{k_1},\overline{\Phi}_{k_2},\Phi_{k_3},\overline{\Phi}_{k_4}\right)\right] \nonumber\\
&=\int \text{d}^2\eta_{\ell_1} \text{d}^2\eta_{\ell_2} \text{d}^2\eta_{k'_1}\text{d}^2\eta_{k'_2}\biggr[\mathcal{A}_3\left(\Phi_{k_1},\overline{\mathcal{W}}_{k'_1},\mathcal{G}_{\ell_1}\right)\mathcal{A}_3\left(\mathcal{W}_{k'_2},\overline{\Phi}_{k_2},\mathcal{G}_{-\ell_2}\right)\nonumber\\
    & \hspace{45mm}\times \mathcal{A}_3\left(\Phi_{k_3},\overline{\mathcal{W}}_{-k'_2},\mathcal{G}_{-\ell_1}\right)\mathcal{A}_3\left(\mathcal{W}_{-k'_1},\overline{\Phi}_{k_4},\mathcal{G}_{\ell_2}\right)\bigg]\nonumber\\
    &= 4g_\text{YM}^4m_1^2 m_2^2\Bigg(\sigma+1+\frac{q^2}{2m_1 m_2}\Bigg).
\end{align}
Let us now use this result to obtain its analog in $\mathcal{N}=8$ supergravity. Here, the $\frac12$-BPS superfield $\mathcal{K}$ is the same as a long $\mathcal{N}=4$ multiplet. Its explicit form can be found in equation (2.15) of \cite{Engelbrecht:2022aao}. The 3-point superamplitude for the interaction of a pair of BPS particles with a graviton superfield  $\mathcal{H}$ is then the square of the $\mathcal{N}=4$ SYM amplitude:
\begin{equation}
    \mathcal{M}_3\left(\mathcal{K}_1,\overline{\mathcal{K}}_2,\mathcal{H}_3\right) = \left[\mathcal{A}_3\left(\mathcal{W}_1,\overline{\mathcal{W}}_2,\mathcal{G}_3\right)\right]^2\,.
\end{equation}
As a result, the 4-particle cuts are also simply related as $\text{Cut}\left(\mathcal{M}_4\right) \propto \left(\text{Cut}\left(\mathcal{A}_4\right)\right)^2$. As always in the double-copy, there is an ambiguous overall (dimensionful) constant of proportionality in the relation between $g_{\text{YM}}^2$ and $\kappa$, we fix this by requiring the tree-level gravitational potential agrees with Newton's law. This can be done in a straightforward manner by projecting the 3-point superamplitudes for (anti-)BPS scalars in gauge theory and gravity onto the top of the massless supermultiplet (i.e. positive helicity gluon and graviton):
\begin{equation}
    \mathcal{A}_3\left( \Phi,\overline{\Phi},g^+ \right) = g_{YM}\frac{m}{x},\quad \mathcal{M}_3\left( \Phi,\overline{\Phi},h^{++} \right) = \frac{\kappa}{2}\frac{m^2}{x^2},
\end{equation}
such that the correspondence reads $g_{YM}^2\to \kappa/2$. All in all, we find that the complete 1-loop amplitude in $\mathcal{N}=8$ SUGRA can be written in terms of the box and cross-box cuts as given \eqref{eq:N8}.

\section{Corrections to the Potential and Energy Levels}
\label{app:quantum energy levels}

In this appendix, we address the correspondence between corrections to the quantum potential and corrections to the energy levels of a bound system. As mentioned in Section \ref{sec:quantum}, it is possible in principle that different contributions to the potential cancel each other out when computing the energy eigenvalues. This is due to the virial theorem which equates $\left\langle \frac{p^2}{\mu} \right\rangle=\left\langle \frac{\alpha}{r} \right\rangle$. However in this appendix, we show that demanding degeneracy of the energy levels of our quantum system with respect to the orbital quantum number $l$ requires that each individual correction to the potential vanishes separately. We start by proving that this must happen order-by-order in Born perturbation theory (PT) and thus specifically at first order. Next, we argue that corrections to the potential at different orders in the velocity (PN) and weak coupling (PM) expansion will always lead to corrections to the energy levels which have independent functional dependence in the large $n,l$ limit and thus must vanish independently if energy level degeneracy is to be protected.

\subsection{Born perturbation theory}

We begin with corrections from different orders in PT and show that they cannot combine together to completely cancel the $l$-dependence. As a result, we only need to worry about computing the first order in the Born expansion. We start with an uncorrected Hamiltonian of the form
\begin{align}
    H= \frac{p^2}{2\mu} -\frac{\alpha}{r}\,.
\end{align}
We then work with arbitrary classical $\oh(\hbar^0)$ and quantum $\oh(\hbar^1)$ perturbations to the potential of the form
\begin{equation}
\label{eq:correctiontypes}
\begin{aligned}
    \delta V^{(c)}_{k,m} &= v^{(c)}_{k,m} \frac{\alpha^m p^{2k}}{\mu^{m+2k-1}c^{2m+2k-2}r^m}\,,\quad  k+m\geq2.\\
    \delta V_{k,m}^{(q)} &= v^{(q)}_{k,m}\frac{\hbar \alpha^m p^{2k}}{\mu^{m+2k}c^{2m+2k-1}r^{m+1}}\,,\quad  m\ge 2\,, k\ge 0\,.
\end{aligned}
\end{equation}
where $\mu$ is the reduced mass of the two-body system, $v^{(i)}$ are parameters that determine the potential and the $(c)$, $(q)$ superscripts indicate classical and quantum corrections, respectively. Using the relation:
\begin{equation}
    p^2 = 2\mu\left( H_0+\frac{\alpha}{r} \right)\,,
\end{equation}
we can express all the powers of momentum in terms of unperturbed energies $E_n^{(0)}$ and extra powers of $\alpha/r$ when evaluated on-shell. The energy corrections take the schematic form\footnote{Note that commutator terms do not change the scaling of the energy corrections. This is because 
\begin{equation}
\label{eq:commutator}
    \left[ H_0, \frac{\alpha}{r} \right] \sim \frac{\hbar^2\alpha}{\mu r^2}\partial_r,
\end{equation}
and therefore we can use integration-by-parts relations and the Schr\"odinger equation to move each derivative to one more power of $1/a_0$. We then have
\begin{equation}
    \left\langle\left[H_0, \frac{\alpha}{r} \right]\right\rangle \sim \frac{\mu^2\alpha^4}{\hbar^4}\sim \left\langle H_0 \frac{\alpha}{r}\right\rangle \,.
\end{equation}
Thus all terms that can appear in the expansion of the powers of momentum lead to the same uniform scaling of dimensionful quantities.}
\begin{equation}
\label{eq: DeltaE1}
    \Delta E^{(1,c)}_{k,m} = \left\langle \delta V^{(c)}_{k,m}\right\rangle \sim \frac{1}{(\mu c^2)^{m+k-1}}\sum_{j=0}^k \alpha^{m+j}\left(E_n^{(0)}\right)^{k-j}\left\langle \frac{1}{r^{m+j}}\right\rangle\,.
\end{equation}
The relevant expectation values in a central potential are given by the Kramers' recursive formula
\begin{align}
    \frac{k+1}{n^2}\langle r^k\rangle - (2k+1) a_0 \langle r^{k{-}1}\rangle +\frac{k}{4} \left((2l+1)^2-k^2\right) a_0^2 \langle r^{k{-}2}\rangle=0\,, a_0=\frac{\hbar^2}{\mu\alpha}\,,
\end{align}
with the seeds 
\begin{align}
    \left<\frac{1}{r}\right>=\frac{1}{a_0 n^2} \,,\hspace{10mm}\left<\frac{1}{r^2}\right>=\frac{1}{a_0^2 (l+\frac12)n^3}\,.
\end{align}
Putting these together we see that $\langle r^k\rangle \sim a_0^k$. Together with $E_n^{(0)}\sim \frac{\alpha}{a_0}$, we see that the scaling of energy corrections must be
\begin{equation}
    \Delta E^{(1,c)}_{k,m}\sim \mu c^2\left(\frac{\alpha}{c\hbar}\right)^{2k+2m}\,.
\end{equation}
For the quantum corrections, we get a similar result
\begin{equation}
    \Delta E^{(1,q)}_{k,m} \sim \mu c^2 \left(\frac{\alpha}{c\hbar}\right)^{2m+2k+1}.
\end{equation}
Since they contribute at even and odd powers of $\alpha$ respectively, this makes clear that classical and quantum corrections cannot mix at first order in PT, as expected.

If we now consider the energy corrections from the second order in the Born expansion, they are defined as
\begin{equation}
    \Delta E^{(2)}_{nl} = \langle n\,l\,m| \delta V\, P_{nl}\,(E^{(0)}_n-H_0)^{-1}P_{nl}\,\delta V|n\,l\,m\rangle,
\end{equation}
where the projector $P_{nl}$ removes off-diagonal elements with the same principal quantum number $n$, since their matrix elements vanish for scalar operators $\delta V$. These corrections can involve cross terms between classical and quantum corrections to the potential. Concretely, for a pair $(\delta V^{(q)}_{k_1,m_1},\delta V^{(c)}_{k_2,m_2})$
\begin{equation}
    \Delta E^{(2)} \sim \mu c^2\left( \frac{\alpha}{c\hbar} \right)^{2(k_1+k_2+m_1+m_2)-1}.
\end{equation}
Naively, it seems that these corrections could mix with the terms $\Delta E^{(1,q)}$ from first order in PT. However, if we take into account the inequalities that $k$ and $m$ must satisfy as given in \eqref{eq:correctiontypes}, we see that $k_i+m_i\geq2$ for both $i=1,2$. Consider $k_i+m_i=2$. Since the second order corrections $\Delta E^{(2)}$ start at order $\alpha^7$, corrections $\Delta E^{(1)}$ of order $\alpha^5$ do not mix with them. For the energy to remain uncorrected, $\Delta E^{(1)}$ must cancel. 

We can now apply this argument recursively. The first non-vanishing term is now $k_1+k_2+m_1+m_2 = 5$, which means the lower powers in $\Delta E^{(1)}$ again have to cancel independently of $\Delta E^{(2)}$. Proceeding in this way, it is clear that imposing that energy levels do not shift at all orders in $\alpha$ has to be satisfied already at first order in PT.

\subsection{Linear independence of the potential corrections}

Now that we have shown that corrections to the energy levels have to vanish at the first order in the Born expansion, we now prove that different operators $\delta V_{k,m}$ lead to linearly independent contributions to $\Delta E^{(1)}$ with respect to the analytic behavior in the quantum numbers $n,l$. Thus they need to be set to zero individually in order for the energy levels to be degenerate.

Consider the strict $n$, $l\to\infty$ limit, which corresponds to the semi-classical regime. We parametrize it by $l=n\,\lambda$, with $0<\lambda<1$ not being parametrically smaller than 1. Now, if exact degeneracy of the energy levels is preserved, it obviously has to hold for this limit as well. In this regime, we can safely ignore commutator terms that arise when computing expectation values of the different operators. Indeed in \eqref{eq:commutator}, we saw that every commutator yields another power of $1/r$. Since the expectation value of the radial coordinate scales as
\begin{equation}
    \langle r^{-q}\rangle \sim \frac{C_q(\lambda)}{n^{2q}}\,,
\end{equation}
where $C_q(\lambda)$ is a rational function of $\lambda$, we get
\begin{equation}
    \left\langle\left[ H_0,\frac{1}{r} \right]\right\rangle \sim \frac{1}{n^6}\,.
\end{equation}
Meanwhile, the term without a commutator scales as
\begin{equation}
    \left\langle H_0\frac{1}{r} \right\rangle \sim \frac{E_n^{(0)}}{n^2} \sim \frac{1}{n^4},
\end{equation}
so it is dominant with respect to the commutator. This still happens for higher order commutators as well, where now second derivatives are eliminated by using the Schr\"odinger equation
\begin{equation}
    \left\langle\left[ H_0,\left[ H_0,\frac{1}{r} \right] \right]\right\rangle \sim \frac{1}{n^8}\,.
\end{equation}
Again, the terms without commutators are dominant
\begin{equation}
    \left\langle H_0^2\frac{1}{r} \right\rangle \sim \frac{\left(E_n^{(0)}\right)^2}{n^2} \sim\frac{1}{n^6}.
\end{equation}
In conclusion, at leading order when $n,l\to\infty$ simultaneously, we don't need to worry about operator ordering in the expectation values.

Let us now reconsider \eqref{eq: DeltaE1} that describes the contribution of some set of operators that could mix together with the first quantum corrections to the 1-loop potential
\begin{equation}
    \delta V_{s-m+2,m}^{(q)} \sim \frac{p^{2(s-m+2)}\alpha^{m}}{r^{1+m}},\quad 2\leq m\leq s+2.
\end{equation}
We now take $n,\,l\to\infty$ and consider the full $\lambda$ dependence of these contributions
\begin{equation}
\begin{aligned}
    \langle\delta V_{s-m+2,m}^{(q)}\rangle\sim\sum_{q=1+m}^{1+s}\binom{s-m+2}{q-1-m}2^{q-m-1}(-1)^{1+s-q}\, C_q(\lambda),
\end{aligned}
\end{equation}
where we have factored out a uniform $n$ scaling. Now, we see that the correspondence between the operators and the set of $C_q$ functions looks like
\begin{equation}
    \begin{aligned}
        \langle\delta V_{s,2}^{(q)}\rangle &\to \{C_3,C_4,\ldots,C_{3+s}\},\\
        \langle\delta V_{s-1,3}^{(q)}\rangle &\to \{C_4,\ldots,C_{3+s}\},\\
        &\vdots\\\langle\delta V_{0,s+2}^{(q)}\rangle &\to \{C_{3+s}\}.
    \end{aligned}
\end{equation}
In other words, if we define a linear map $\langle\mathbf{\delta V}\rangle = \mathbf{M}\times\mathbf{C}$, where $\mathbf{M}$ is a $(s+1)\times (s+1)$ matrix and $\langle\mathbf{\delta V}\rangle,\mathbf{C}$ are $(s+1)$-component vectors, then $\mathbf{M}$ is upper triangular. In addition, since the diagonal elements ($q=1+s$) are always non-zero
\begin{equation}
    M_{i,i} = (-1)^{s+1-i}\neq0,
\end{equation}
we conclude that $\mathbf{M}$ is always invertible. Therefore, the set of operators $\delta V_{s-m+2,m}$ (at fixed $r$) spans the same vector space as the set of functions $C_q(\lambda)$. We now just have to prove that the latter are linearly independent with respect to the dependence on the orbital quantum number $l$.

To do that, we use Kramers' recursion relation in the large $(n,l)$ limit
\begin{equation}
    C_q(\lambda) = \frac{1}{\lambda^2(q-2)}\Big[(2q-5)C_{q-1}(\lambda) - (q-3)C_{q-2}(\lambda)\Big]. 
\end{equation}
Taking into account that $C_1(\lambda)=1,\ C_2(\lambda) = \lambda^{-1}$, at each order $q$ the dominant pole will be $\lambda^{3-2q}$, which makes it impossible to construct a linear combination of functions $C_q(\lambda)$ that completely cancels out the dependence on $\lambda$. Thus, we conclude that in order to cancel the $l$-dependence in $\Delta E^{(1)}$ for the bound states, we need each term in the potential to vanish individually.

\section{Details of the Gravitino Exchange Calculation}
\label{app:gravitino details}

In this appendix we provide further details of the calculation described in Section \ref{subsec:gravitino} of the contribution to the quantum 2PM potential due to the exchange of a pair of massless spin-3/2 gravitinos. As described, the 1-loop integrand is constructed by sewing the cut depicted in Figure \ref{fig:t-channel-cut}. The Compton amplitudes we require in this case have a pair of external massive scalars $\overline{\Phi}, \Phi$ and a pair of external massless gravitinos $\overline{\psi}, \psi$. Without loss of generality, it is convenient in this calculation to treat the gravitinos as Dirac, the number of states in the cut is then related as $n_{3/2}^{\text{Dirac}} = 2 n_{3/2}^{\text{Majorana}}$. 

The gravitino Compton amplitude can be efficiently constructed as a BCJ double-copy \cite{Kawai:1985xq,Bern:2008qj}. The ``single-copy'' model we consider is Yang-Mills coupled to a collection of complex/Dirac matter fields in the adjoint representation: a scalar $\Phi$ and spin-1/2 fermion $\Psi$ of mass $m$ and a massless spin-1/2 fermion $\chi$. In addition to the usual gauge coupling $g_{\text{YM}}$, the model has a Yukawa coupling $\mathcal{L}\supset g_{\text{Y}} \text{Tr}\left[\overline{\Phi} \overline{\chi}\Psi + \text{h.c.}\right]$. In BCJ form the ``left'' Compton amplitude we need is
\begin{equation}
    \mathcal{A}^{\text{L}}_4\left(\overline{\Phi}_{p_1}, g_{p_2}, g_{p_3}, \Phi_{p_4}\right) = \frac{c_{12} n^{\text{L}}_{12}}{s_{12}-m^2} + \frac{c_{13} n^{\text{L}}_{13}}{s_{13}-m^2} + \frac{c_{23} n^{\text{L}}_{23}}{s_{23}},
\end{equation}
where the numerators
\begin{align}
    \label{eq:numeratorsL}
    n^{\text{L}}_{12} &=  -2g_{\text{YM}}^2\left[(p_1\cdot p_2)(\varepsilon_2 \cdot \varepsilon_3) - 2(p_1 \cdot \varepsilon_2)(p_1\cdot \varepsilon_3)- 2(p_1 \cdot \varepsilon_2)(p_2\cdot \varepsilon_3)\right] \nonumber\\
    n^{\text{L}}_{13} &=  2g_{\text{YM}}^2\left[(p_1\cdot p_3)(\varepsilon_2 \cdot \varepsilon_3) - 2(p_1 \cdot \varepsilon_3)(p_1\cdot \varepsilon_2)- 2(p_1 \cdot \varepsilon_3)(p_3\cdot \varepsilon_2)\right]\nonumber\\
    n^{\text{L}}_{23} &= -n^{\text{L}}_{12} - n^{\text{L}}_{13},
\end{align}
are manifestly color-kinematics dual.\footnote{The color factors are defined as $c_{12}:=f^{a_1 a_2 b} f^{a_3 a_4 b}$, $c_{13}:=f^{a_1 a_3 b} f^{a_4 a_2 b}$, $c_{23}:=f^{a_1 a_4 b} f^{a_2 a_3 b}$ and satisfy the Jacobi identity $c_{12}+c_{13}+c_{23}=0$.} Similarly the ``right'' Compton amplitude is given by 
\begin{equation}
    \mathcal{A}^{\text{R}}_4\left(\overline{\Phi}_{p_1}, \overline{\chi}_{p_2}, \chi_{p_3}, \Phi_{p_4}\right) = \frac{c_{12} n^{\text{R}}_{12}}{s_{12}-m^2} + \frac{c_{13} n^{\text{R}}_{13}}{s_{13}-m^2} + \frac{c_{23} n^{\text{R}}_{23}}{s_{23}},
\end{equation}
where
\begin{align}
    \label{eq:numeratorsR}
    n^{\text{R}}_{12} &= -g_{\text{Y}}^2 \overline{u}(p_2)\left(\slashed{p}_1 +m\right)v(p_3)\nonumber\\
    n^{\text{R}}_{13} &= -g_{\text{Y}}^2 \overline{u}(p_2)\left(\slashed{p}_1 -m\right)v(p_3)\nonumber\\
    n^{\text{R}}_{23} &= g_{\text{YM}}^2 \overline{u}(p_2)\slashed{p}_1 v(p_3).
\end{align}
By contrast these numerators are not color-kinematics dual for generic values of the Yukawa coupling; imposing the condition $n^R_{12}+n^R_{13}+n^R_{23}=0$ requires $2g_{\text{Y}}^2 \overset{!}{=} g_{\text{YM}}^2$. The double-copy amplitude is then given by
\begin{equation}
    \mathcal{M}^{\text{tree}}_4\left(\overline{\Phi}_{p_1}, \overline{\psi}_{p_2}, \psi_{p_3}, \Phi_{p_4}\right) = \frac{n^{\text{L}}_{12} n^{\text{R}}_{12}}{s_{12}-m^2} + \frac{n^{\text{L}}_{13} n^{\text{R}}_{13}}{s_{13}-m^2} + \frac{n^{\text{L}}_{23} n^{\text{R}}_{23}}{s_{23}},
\end{equation}
where the map between the gauge and gravitational couplings is explicitly
\begin{equation}
    g_{\text{YM}}^2 \mapsto -\frac{\kappa^2}{8}.
\end{equation}
The precise proportionality coefficient can be fixed by comparing with soft gravitino limits, which are sensitive only to 3-particle interactions at leading-order. The relative coefficients of the 3-particle couplings $h\Phi \Phi$, $hhh$ and $h\psi \psi$ are in-turn fixed by imposing the mutual consistency of the 4-particle soft limit Ward identities (Weinberg soft constraints) \cite{Weinberg:1964ew}. In this covariant form the external gravitino polarization vector-spinors are identified as 
\begin{equation}
    \varepsilon^\mu_a(p) \sim \varepsilon^\mu(p) u_a(p),\hspace{10mm} (\gamma^\mu)_{ab}  \varepsilon_\mu(p) u_b(p) \overset{!}{=}0,
\end{equation}
where $a,b$ are 4-component spinor indices. The second condition imposes the on-shell ``gamma-traceless'' condition and projects out the lower spin component of the tensor product $1\otimes \frac{1}{2} = \frac{3}{2} \oplus \frac{1}{2}$. 

To calculate the long-range part of the 1-loop integrand we sew the cut depicted in Figure \ref{fig:t-channel-cut}
\begin{align}
    &\text{Cut}_{\bub{}}\left[i\mathcal{M}^{\text{1-loop}}_4\left(\overline{\Phi}_{-p_1},\Phi_{p_1+q},\overline{\Phi}_{-p_2},\Phi_{p_2-q}\right)\right] \nonumber\\
    &= -\text{Tr}\left[\left(i\slashed{\mathcal{M}}^{\text{tree}}_4\left(\overline{\Phi}_{-p_1}, \overline{\psi}^{\mu_1}_{\ell_1}, \psi^{\nu_2}_{-\ell_2}, \Phi_{p_1+q}\right)\right)\cdot\left(i\slashed{\Pi}_{\frac{3}{2}}(\ell_2)_{\nu_2 \mu_2}\right)\right.\nonumber\\
    &\hspace{15mm} \left.\cdot\left(i\slashed{\mathcal{M}}^{\text{tree}}_4\left(\overline{\Phi}_{-p_2}, \overline{\psi}^{\mu_2}_{\ell_2}, \psi^{\nu_1}_{-\ell_1}, \Phi_{p_2-q}\right)\right)\cdot\left(i\slashed{\Pi}_{\frac{3}{2}}(\ell_1)_{\nu_1 \mu_1}\right)\right],
\end{align}
where there is an overall minus sign in front of the Dirac trace since we are calculating a fermion loop. The massless spin-3/2 projectors, in light-cone gauge are given explicitly as 
\begin{equation}
    \slashed{\Pi}_{\frac{3}{2}}(q,n)^{\mu \nu} := \slashed{q}\left[\Pi_1^{\mu\nu}(q,n) - \frac{1}{2} \Pi_1^{\mu\rho}(q,n) \gamma_\rho \gamma_\sigma \Pi_1^{\sigma\nu}(q,n)\right],
\end{equation}
with the spin-1 light-cone projector
\begin{equation}
    \Pi_1^{\mu\nu}(q,n) := \eta^{\mu\nu} - \frac{q^\mu n^\nu + n^\mu q^\nu}{q\cdot n},
\end{equation}
where $n^\mu$ is an auxiliary light-like vector $n^2=0$, with $n\cdot q\neq 0$. Above we implicitly define the stripped Compton amplitude $\slashed{\mathcal{M}}_4^{\text{tree}}\left(\overline{\Phi}_1,\overline\psi^\mu_2,\psi^\nu_3, \Phi_4\right)$ by removing the polarization vector-spinors of the gravitinos. Using the form of the numerators (\ref{eq:numeratorsL}) and (\ref{eq:numeratorsR}) verbatim gives an expression for this stripped amplitude that satisfies a strong version of the generalized Ward identity \cite{Kosmopoulos:2020pcd}
\begin{equation}
p_{2\mu}\slashed{\mathcal{M}}_4^{\text{tree}}\left(\overline{\Phi}_1,\overline\psi^\mu_2,\psi^\nu_3, \Phi_4\right) = p_{3\nu}\slashed{\mathcal{M}}_4^{\text{tree}}\left(\overline{\Phi}_1,\overline\psi^\mu_2,\psi^\nu_3, \Phi_4\right) = 0.
\end{equation}
We can therefore use a simplified ``Feynman gauge'' version of the spin-3/2 projectors in the cut sewing
\begin{equation}
    \slashed{\Pi}_{\frac{3}{2}}(q)^{\mu \nu} \cong \slashed{q}\left[\eta^{\mu\nu} - \frac{1}{2} \gamma^\mu \gamma^\nu \right].
\end{equation}
Sewing the gravitino cut, expanding to quantum order and calculating the soft integrals as described in Appendix \ref{app:integrals} gives the contribution of a single \textit{Majorana} gravitino to the amplitude (\ref{eq:gravitino amplitude}).

\bibliographystyle{utphys}
\bibliography{newbib}

\end{document}